\documentclass[aps,pre,nofootinbib,twocolumn,floatfix,superscriptaddress,showpacs,longbibliography]{revtex4-1}

\usepackage{soul}
\usepackage{amsfonts}
\usepackage{amsmath}
\usepackage{amssymb}
\usepackage{mathtools}
\usepackage{graphicx}
\usepackage{xcolor}
\usepackage{bbm}
\usepackage[colorlinks=true,linkcolor=blue,citecolor=blue,urlcolor=blue]{hyperref}
\usepackage{calrsfs}
\DeclareMathAlphabet{\pazocal}{OMS}{zplm}{m}{n}
\usepackage{subfigure}
\usepackage{mathptmx}
\usepackage{times,txfonts}
\usepackage{siunitx}
\usepackage{csquotes}
\usepackage{enumerate}

\newcommand{\tr}{{\rm tr}}
\newcommand{\heat}[1]{\dot{\mathcal{Q}}_{#1}}
\newcommand{\pow}{\mathcal{P}}
\newcommand{\nobibentry}[1]{{\let\nocite\ignore\bibentry{#1}}}

\newcommand{\bibfnamefont}[1]{#1}
\newcommand{\bibnamefont}[1]{#1}

\newcommand{\ket}[1]{\left\vert#1\right\rangle}
\newcommand{\bra}[1]{\left\langle#1\right\vert}
\newcommand*\diff{\mathop{}\!\mathrm{d}}

\renewcommand\Re{\operatorname{Re}}
\renewcommand\Im{\operatorname{Im}}

\definecolor{myblue}{HTML}{000000}

\begin{document} 

\title{Classical emulation of quantum-coherent thermal machines}

\author{J. Onam Gonz{\'a}lez} 
\email{jgonzall@ull.es}
\affiliation{Departamento de F{\'i}sica, Universidad de La Laguna, La Laguna 38204, Spain}
\affiliation{IUdEA, Universidad de La Laguna, La Laguna 38204, Spain}

\author{Jos{\'e} P. Palao}
\email{jppalao@ull.edu.es}
\affiliation{Departamento de F{\'i}sica, Universidad de La Laguna, La Laguna 38204, Spain}
\affiliation{IUdEA, Universidad de La Laguna, La Laguna 38204, Spain}

\author{Daniel Alonso}
\email{dalonso@ull.edu.es}
\affiliation{Departamento de F{\'i}sica, Universidad de La Laguna, La Laguna 38204, Spain}
\affiliation{IUdEA, Universidad de La Laguna, La Laguna 38204, Spain}

\author{Luis A. Correa} 
\email{luis.correa@nottingham.ac.uk}
\affiliation{School of Mathematical Sciences and CQNE, The University of Nottingham, University Park, Nottingham NG7 2RD, United Kingdom}
\affiliation{Kavli Institute for Theoretical Physics University of California, Santa Barbara, CA 93106}

\begin{abstract}
The performance enhancements observed in various models of continuous quantum thermal machines have been linked to the buildup of coherences in a preferred basis. But, is this connection always an evidence of `quantum-thermodynamic supremacy'? By force of example, we show that this is not the case. In particular, we compare a power-driven three-level {\color{myblue}continuous} quantum refrigerator with a four-level combined cycle, partly driven by power and partly by heat. We focus on the weak driving regime and find the four-level model to be superior since it can operate in parameter regimes in which the three-level model cannot, it may exhibit a larger cooling rate, and, simultaneously, a better coefficient of performance. Furthermore, we find that the improvement in the cooling rate matches the increase in the stationary quantum coherences \textit{exactly}. Crucially, though, we also show that the thermodynamic variables for both models follow from a classical representation based on graph theory. This implies that we can build incoherent stochastic-thermodynamic models with the same steady-state operation or, equivalently, that both coherent refrigerators can be emulated classically. More generally, we prove this for \textit{any} $ N $-level weakly driven device with a `cyclic' pattern of transitions. {\color{myblue}Therefore, even if coherence is present in a specific quantum thermal machine, it is often not essential to replicate the underlying energy conversion process}.
\end{abstract}

\date{\today}

\maketitle

\section{Introduction} \label{sec:introduction}

`Quantum thermodynamics' studies the emergence of thermodynamic behaviour in individual quantum systems \cite{binder2018thermodynamics}. Over the past few years, the field has developed very rapidly \cite{Kosloff2013,Kosloff2014,gelbwaser2015thermodynamics,goold2016role,Vinjanampathy2016} and yet, key recurring questions remain unanswered:
\begin{displayquote}
What is \emph{quantum} in quantum thermodynamics?
\end{displayquote}
\begin{displayquote}
Can quantum heat devices exploit \textit{quantumness} to outperform their classical counterparts?
\end{displayquote}

Quantum thermal machines are the workhorse of quantum thermodynamics. Very generally, these consist of an individual system $ S $ which can couple to various heat baths at different temperatures and, possibly, is also subject to dynamical control by an external field. After a transient, $ S $ reaches a non-equilibrium steady state characterized by certain rates of energy exchange with the heat baths. The direction of these energy fluxes can be chosen by engineering $ S $, which may result in, e.g., a heat engine \cite{PhysRevLett.2.262} or a refrigerator \cite{5123190}. Considerable efforts have been devoted to optimize these devices \cite{Geva1996,Feldmann2000,felmann2006lubrication,rezek2006irreversible,Esposito2009,Creatore2013,Correa2014,Correa2014b,Correa2014c,Gelbwaser2015,Niedenzu2015,PhysRevE.92.032136,palao2016efficiency,Uzdin2016} and to understand whether genuinely \textit{quantum} features play an active role in their operation \cite{Scully2003,Scully2010,Scully2011,Svidzinsky2011,feldmann2012short,Correa2014a,brunner2014entanglement,Uzdin2015,
Killoran2015,brask2015small,mitchison2015coherence,Xu2016,Silva2016,Friedenberger2017a,Grangier2018,Dorfman2018,
Houlebec2018,Du2018,Kilgour2018,Gelbwaser2018}. 

One might say that a thermal machine is \textit{quantum} provided that $ S $ has a discrete spectrum. In fact, the energy filtering allowed by such discreteness can be said to be advantageous, since it enables continuous energy conversion at the (reversible) Carnot limit of maximum efficiency \cite{PhysRev.156.343,Esposito2009,Correa2014b}. Similarly, energy quantization in multi-stroke thermodynamic cycles can give rise to experimentally testable non-classical effects \cite{Gelbwaser2018}. In most cases, however, it is attributes such as entanglement or coherence which are regarded as the hallmark of genuine quantumness. 

In particular, quantum coherence \cite{Baumgratz2014,Streltsov2017} has often been seen as a potential resource, since it can influence {\color{myblue}the thermodynamically-relevant quantities, such heat and work, of open systems} \cite{Bulnes2016}. It has been argued, for instance, that radiatively and noise-induced coherences \cite{Scully2003,Scully2010,Scully2011,Svidzinsky2011} might enhance the operation of quantum heat engines \cite{Killoran2015,Xu2016,Dorfman2018} and heat-driven quantum refrigerators \cite{Houlebec2018,Du2018,Kilgour2018}. However, it is not clear whether they are truly \textit{instrumental} \cite{Gelbwaser2015,Niedenzu2015,Houlebec2018,Kilgour2018}, since similar effects can be obtained from stochastic-thermodynamic models \cite{Seifert2012,Broeck2015,Polettini2016,nimmrichter2017quantum}, i.e., classical incoherent systems whose dynamics is governed by balance equations concerning \textit{only} the populations in some relevant basis (usually, the energy basis).

A possible approach to elucidate the role of quantum coherence in any given model is to add dephasing, thus making it fully incoherent (or classical) \cite{Uzdin2015,Uzdin2016,Kilgour2018}. An ensuing reduction in performance would be an evidence of the usefulness of coherence in quantum thermodynamics. Furthermore, if the ultimate limits on the performance of incoherent thermal machines can be established, coherences would become \textit{thermodynamically} detectable---one would simply need to search for violations of such bounds \cite{Uzdin2016,klatzow2017experimental}.

In this paper we adopt a much more stringent operational definition for `quantumness': \textit{No thermal machine should be classified as quantum if its  thermodynamically-relevant quantities can be replicated exactly by an incoherent emulator\footnote{Note that throughout this paper, we only require the emulator to replicate the \textit{averaged} heat flows, but we make not mention to their fluctuations. The emulability of higher-order moments of the fluxes is an interesting point that certainly deserves separate analysis.}}. More precisely, the emulator should be a classical dissipative system operating between the same heat baths and with the same {\color{myblue}frequency gaps} and number of discrete states. Interestingly, {\color{myblue} we will show that the currents of many continuous quantum-coherent devices are thermodynamically indistinguishable from those of their `classical emulators'}.

If it exists, such emulator needs not be related to the coherent device of interest by the mere addition of dephasing---it can be a different model so long as it remains incoherent at all times and that, once in the stationary regime, it exchanges energy with its surroundings at the same rates as the original machine. In particular, the transient dynamics of the coherent model can be very different from that of its emulator. In the steady state, however, it must be impossible to tell one from the other by only looking at heat fluxes and power. 

For simplicity, we focus on periodically driven continuous refrigerators with a `cyclic' scheme of transitions (see Fig.~\ref{fig:fig1}), although, as we shall point out, our results apply to devices with more complex transition patterns. Specifically, `continuous' thermal machines \cite{Kosloff2014} are models in which the working substance $ S $ couples \textit{simultaneously} to a cold bath at temperature $ T_c $, a hot bath at $ T_h > T_c $, and a classical field \footnote{In an absorption refrigerator the driving is replaced by thermal coupling to a `work bath' at temperature $ T_w > T_h $, which drives the cooling process.}. Since the driving field is periodic, we must think of the steady state of the machine as a `limit cycle' where all thermodynamic variables are evaluated as time averages. Concretely, a quantum refrigerator can drive heat transport against the temperature gradient in a suitable parameter range, or `cooling window'. We work in the limit of \textit{very weak} driving, which allows us to derive a `local' master equation for $ S $ \cite{gonzalez2017testing}. We show that stationary quantum coherence is not only present in all these models but, in fact, it is \textit{essential} for the energy-conversion process to take place. Strikingly, however, our main result is that the steady-state operation of any such quantum-coherent $ N $-level machine admits a classical representation based on graph theory \cite{Hill1966,Schnakenberg1976,Polettini2016,Gonzalez2016,Gonzalez2017}. It follows that an incoherent device can always be built such that its steady-state thermodynamic variables coincide with those of the original model. Hence, {\color{myblue} this} entire family of quantum-coherent thermal machines can be emulated classically. The design of such emulator is reminiscent of the mapping of the limit cycle of a periodically driven classical system to {\color{myblue} a} nonequilibrium steady state \cite{Raz2016}. {\color{myblue} We want to stress that, from now on, we focus exclusively on the steady-state operation of continuous quantum heat devices. In particular, this leaves out all `reciprocating' machines. We note, however, that the latter reduce to the former in the limit of `weak action' \cite{Uzdin2015}.}

As an illustration, we consider the paradigmatic power-driven three-level refrigerator \cite{5123190,Geva96,Palao2001}, which we use as a benchmark for a novel four-level hybrid device, driven by a mixture of heat and work. Concretely, we show that our new model may have a wider cooling window, larger cooling power, and larger coefficient of performance. We also show that the energy-conversion rate in both models is proportional to their steady state coherence. As a result, the excess coherence of the four-level model relative to the benchmark matches exactly the cooling enhancement. It would thus seem that quantum coherence is necessary for continuous refrigeration in the weak driving limit and that the improved cooling performance of the four-level model can be fully attributed to its larger steady-state coherence. If so, observing a non-vanishing `cooling rate' in either device, or certifying that the cooling rate of the four-level model is indeed larger than that of the benchmark would be unmistakable signatures of \textit{quantumness}. Crucially, both coherent devices are cyclic and weakly driven and, as such, they cannot be distinguished from their \textit{classical} analogues in a black-box scenario. Therefore, quantum features might not only be present, but even be intimately related to the thermodynamic variables of quantum thermal machines under study and still, there may be nothing necessarily quantum about their operation. We remark, however, that the thermodynamic equivalence between the continuous thermal machine and its emulator only holds in the steady state. Importantly, we shall also see that the graph theory analysis is a convenient and powerful tool \cite{Gonzalez2016,Gonzalez2017} to obtain accurate approximations for the non-trivial steady-state heat currents of these devices. 

The paper is organized as follows: In Sec.~\ref{sec:refrigerators} we introduce our central model of weakly and periodically driven $ N $-level `cyclic' refrigerator. Its steady-state classical emulator is constructed in Sec.~\ref{sec:counterpart}. Some basic concepts of graph theory {\color{myblue}are} also introduced at this point. In particular, we show that the emulator is a single-circuit graph whose heat currents, power, and coefficient of performance may be obtained in a thermodynamically consistent way. The generalization to more complex transition schemes is also discussed at this point. Using the graph-theoretical toolbox, we then analyze, in Sec.~\ref{sec:3&4leve}, our novel four-level device and the three-level benchmark. We thus arrive to analytical expressions indicating improvements in the steady state functioning of the four-level model in a suitable regime. Finally, in Sec.~\ref{sec:conclusions}, we discuss the implications of our results, summarize, and draw our conclusions. 

\section{Cyclic thermal machines} \label{sec:refrigerators}

\subsection{The system Hamiltonian} \label{sec:cyclic_hamiltonian}

\begin{figure}[t!]
\includegraphics[width=\linewidth]{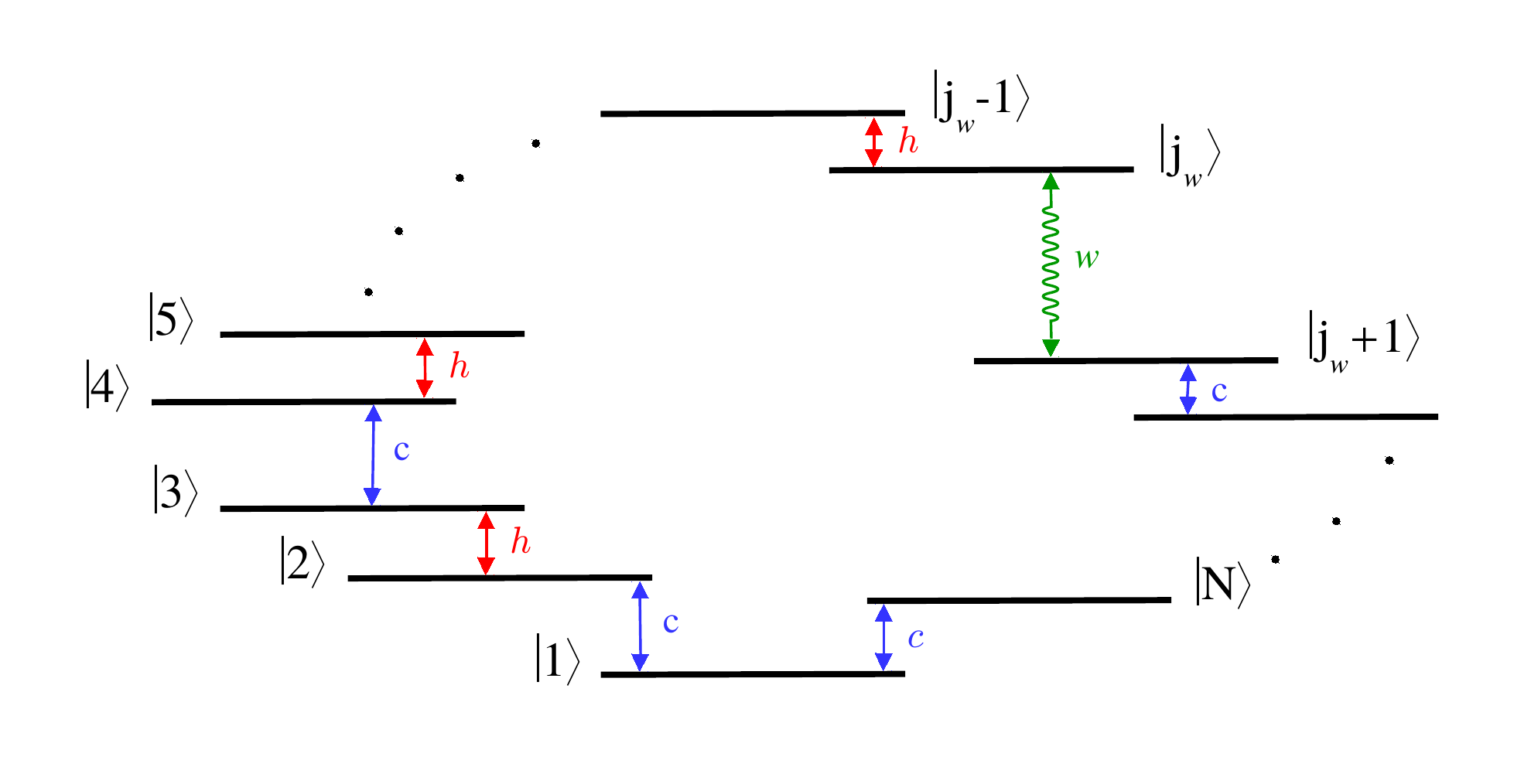}
\caption{(colour online) Energy level diagram for a generic $ N $-level cyclic thermal machine. The labelled red and blue arrows stands for transitions mediated by dissipative interactions with the hot and cold bath, respectively. Their distribution is arbitrary. The periodic field coupling energy levels $\ket{j_w}$ and $\ket{j_w+1}$ is indicated by the wobbly green arrow.} \label{fig:fig1}
\end{figure}

We start by introducing the general model for a coherent cyclic thermal machine (see Fig.~\ref{fig:fig1}). The Hamiltonian for the system or \textit{working substance} $ S $ is comprised of two terms: a bare (time-independent) Hamiltonian $ \hat{H}_\text{0} $ and a time-dependent contribution $ \hat{H}_\text{d}(t) $ which describes the coupling to a sinusoidal driving field. That is,
\begin{subequations}
\begin{eqnarray}
\hat{H}_\text{s}(t) &=& \hat{H}_\text{0} + \hat{H}_\text{d}(t) , \label{eq:HS} \\
\hat{H}_\text{0} &=& \sum\nolimits_{i=1}^{N} E_i\,|i\rangle\langle i| , \label{eq:H0} \\
\hat{H}_\text{d}(t) &=& 2\hbar \lambda \ket{j_w} \bra{j_w+1} \cos{\omega_{j_w}\,t} + \text{h.c.} , \label{eq:HSSw}
\end{eqnarray}
\label{eq:Hs_full}
\end{subequations}
where $ \hbar $ is the reduced Planck constant and $ \lambda $ controls the strength of the interaction with the field. $ E_i $ and $ \ket{i} $ are, respectively, the energies and eigenstates of the bare Hamiltonian. In particular, the driving connects the bare energy states $\ket{j_w} $ and $ \ket{j_w+1} $. For simplicity, we assume a \textit{resonant} coupling, i.e., $\omega_{j_w} \coloneqq (E_{j_w+1}-E_{j_w})/\hbar $, which is optimal from a thermodynamic viewpoint. The generalization to the non-resonant case is, nevertheless, straightforward.

The hot and cold bath can be cast as infinite collections of independent bosonic modes with a well-defined temperature. Their Hamiltonians read
\begin{equation} \label{eq:HBa}
\hat{H}_\alpha = \hbar \sum\nolimits_{\mu} \omega_{\mu,\alpha}\, \hat{b}_{\mu,\alpha}^\dagger \hat{b}_{\mu,\alpha}, \qquad \alpha\in\{c,h\},
\end{equation}
with $ \hat{b}_{\mu,\alpha}^\dagger $ and $ \hat{b}_{\mu,\alpha} $ being the bosonic creation and annihilation operators of the mode at frequency $ \omega_{\mu,\alpha} $ in bath $ \alpha \in \{c,h\} $. For the system-baths couplings, we adopt the general form $ \hat{H}_\text{s--$\alpha$} = \hat{X}_\alpha \otimes \hat{B}_\alpha $, where  
\begin{subequations}
\begin{eqnarray}
\hat{X}_{\alpha} &=& \sum\nolimits_{i\in \mathbf{R}_\alpha} \ket{i}\bra{i+1} + \text{h.c.}, \label{eq:A} \\
\hat{B}_\alpha &=& \hbar \sqrt{\gamma_\alpha} \sum\nolimits_\mu g_{\mu,\alpha}\, \left(\hat{b}_{\mu,\alpha}+\hat{b}_{\mu,\alpha}^\dagger\right). \label{eq:Bf} 
\end{eqnarray}
\end{subequations}
Here, $ g_{\mu,\alpha} \propto \sqrt{\omega_{\mu,\alpha}} $ and $ \gamma_\alpha $ is the \textit{dissipation rate} for bath $ \alpha $. $ \mathbf{R}_\alpha $ stands for the labels of the eigenstates $ \ket{i} $ dissipatively coupled to $ \ket{i+1} $ through the interaction with bath $ \alpha $. For instance, in Fig.~\ref{fig:fig1}, $ \mathbf{R}_c = \{1, 3,\hdots, j_w+1,\hdots, N\} $ and $ \mathbf{R}_h = \{2, 4,\hdots, j_w-1,\hdots\} $. Notice that \textit{all} levels $ \ket{i} $ are thermally coupled to $ \ket{i+1} $ (provided that $ i \neq j_w $) via either the hot or the cold bath. In particular, the $ N $-th level couples to $ i = 1 $, hence closing the cycle. Without loss of generality, we consider that all transitions related to the same bath have different energy gaps, i.e., $ |\omega_k| \neq |\omega_l| $ for $ k,\, l \in \mathbf{R}_\alpha $ ($ k\neq l $). This technical assumption simplifies the master equation but does not restrict the physics of the problem. The full Hamiltonian of the setup is thus
\begin{equation}
\hat{H} = \hat{H}_\text{s} + \sum\nolimits_\alpha\hat{H}_\text{s--$ \alpha $} + \hat{H}_\alpha.
\label{eq:H_total}
\end{equation}

{\color{myblue} Although we consider a cyclic scheme of transitions, our formalism applies to more general quantum-coherent heat devices. Namely, equivalent results can be easily found for non-degenerate systems with various non-consecutive driven transitions. Likewise, one could include parasitic loops to the design. As an illustration, we analyze a model including an extra hot transition between $\ket{j_w}$ and $\ket{j_w+1}$ in the \hyperref[Sec:appendix]{Appendix} below.}

\subsection{The \emph{local} master equation for weak driving}

When deriving an effective equation of motion for the system, it is important to consider the various time scales involved \cite{Breuer2002}. Namely, the bath correlation time $ \tau_\text{B} $, the intrinsic time scale of the bare system $ \tau_\text{0} $, the relaxation time scale $ \tau_\text{R} $, and the typical time associated with the interaction of the bare system with the external field $ \tau_\text{s--d} $. These are
\begin{subequations}
\begin{align}
\tau_\text{B} &\simeq \max\big\lbrace\, \hbar/(k_B T_\text{c}),\,\hbar/(k_B T_\text{h})\, \big\rbrace = \hbar/(k_B T_\text{c}).\\    
\tau_\text{0} &\simeq \max \big\lbrace\, \vert \pm \omega_k \mp \omega_l \vert^{-1},\vert 2\,\omega_k\vert^{-1}\big\rbrace, \qquad (k \neq l). \\
\tau_\text{R} &\simeq \gamma_\alpha^{-1} , \qquad \alpha\in\{c,h\}. \\
\tau_\text{s--d} &\simeq \lambda^{-1}.
\end{align}
\label{eq:time_scales}
\end{subequations}

For a moment, let us switch off the time-dependent term $ \hat{H}_\text{d}(t) $ and discuss the usual weak-coupling Markovian master equation, i.e., the Gorini--Kossakowski--Lindblad--Sudarshan (GKLS) equation \cite{gorini1976completely,lindblad1976generators}. Its microscopic derivation relies on the Born-Markov and secular approximations, which hold whenever $ \tau_{\text{B}} \ll \tau_{\text{R}} $ and $ \tau_{\text{0}} \ll \tau_{\text{R}} $. It can be written as 
\begin{equation} \label{eq:LME}
\frac{\diff \hat{\rho}_\text{s}}{\diff t} = -\frac{\mathrm{i}}{\hbar}[\hat{H}_\text{0},\,\hat{\rho}_\text{s}]+\left(\pazocal{L}_\text{c} + \pazocal{L}_\text{h}\right)\,\hat{\rho}_\text{s},
\end{equation}
where $ \hat{\rho}_\text{s} $ is the reduced state of the $ N $-level system. Crucially, due to the underlying Born approximation of weak dissipation, Eq.~\eqref{eq:LME} is correct only to $ \pazocal{O}\big(\max\{\gamma_\text{c},\gamma_\text{h}\}\big) $. 

The action of the super-operator $ \pazocal{L}_\alpha $ is given by
\begin{multline} \label{eq:dissipator}
\pazocal{L}_\alpha\,\hat{\rho}_\text{s} = \sum\nolimits_{i \in \mathbf{R}_\alpha}  \Gamma_{\omega_i}^{\alpha} \, \bigg(\hat{A}_i\,\hat{\rho}_\text{s}\hat{A}_i^\dagger - \frac12\left\lbrace\hat{A}_i^\dagger\hat{A}_i,\, \hat{\rho}_\text{s}\right\rbrace_+ \bigg) \\ 
+ \Gamma_{-\omega_i}^{\alpha} \, \bigg( \hat{A}_i^\dagger \hat{\rho}_S \hat{A}_i - \frac12\left\lbrace \hat{A}_i\,\hat{A}_i^\dagger,\, \hat{\rho}_\text{s}\right\rbrace_+\bigg). 
\end{multline}
Here, $ \hat{A}_i = \ket{i}\bra{i+1} $ and the notation $ \{\cdot,\cdot\}_+ $ stands for anti-commutator. The `jump' operators $ \hat{A}_i $ are such that $ \hat{X}_\alpha = \sum\nolimits_{i\in\mathbf{R}_\alpha} \big(\hat{A}_i + \hat{A}_i^\dagger\big) $ and $ [\hat{H}_\text{0},\hat{A}_i] = -\omega_i \hat{A}_i $. As a result, the operators $ \hat{X}_\alpha $ in the interaction picture with respect to $ \hat{H}_\text{0} $ read
\begin{equation}
e^{\mathrm{i}\,\hat{H}_\text{0}\,t/\hbar}\,\hat{X}_\alpha\,e^{-\mathrm{i}\,\hat{H}_\text{0}\,t/\hbar} = \sum\nolimits_{i\in\mathbf{R}_\alpha} e^{-\mathrm{i}\,\omega_i\,t}\hat{A}_i.
\label{eq:X_interaction}
\end{equation}
This identity is a key step in the derivation of Eq.~\eqref{eq:LME} \cite{Breuer2002}.

If we now switch $ \hat{H}_\text{d}(t) $ back on, we will need to change the propagator in Eq.~\eqref{eq:X_interaction} over to the time-ordered exponential
\begin{equation}
\hat{U}_\text{s}(t) = \pazocal{T}\,\exp{\left(-\mathrm{i}\,\hbar^{-1}\int_0^t \diff t' \, [ \, \hat{H}_\text{0} + \hat{H}_\text{d}(t') \, ] \, \right)}.
\label{eq:propagator}    
\end{equation}
When deriving a GKLS master equation for such a periodically driven system one also looks for a different decomposition on the right-hand side of Eq.~\eqref{eq:X_interaction} \cite{1205.4552v1,szczygielski2014application,Correa2014b}. Namely,
\begin{equation}
\hat{U}_\text{s}^\dagger(t)\,\hat{X}_\alpha\,\hat{U}_\text{s}(t) = \sum\nolimits_{q\in\mathbb{Z}}\sum\nolimits_{\{\bar{\omega}\}} e^{-\mathrm{i}\,(\bar{\omega}+q\,\omega_{j_w})\,t}\hat{A}_i^{(q)}.
\label{eq:X_Floquet}
\end{equation}
We will skip all the technical details and limit ourselves to note that if, in addition to $ \tau_\text{B} \ll \tau_\text{R} $ and $ \tau_\text{0} \ll \tau_\text{R} $, we make the \textit{weak driving} assumption of $ \tau_\text{R} \ll \tau_\text{s--d} $, Eq.~\eqref{eq:X_Floquet} can be cast as
\begin{align}
\hat{U}_\text{s}^\dagger(t)\,\hat{X}_\alpha\,\hat{U}_\text{s}(t) &= e^{\mathrm{i}\,\hat{H}_\text{0}\,t/\hbar}\,\hat{X}_\alpha\,e^{-\mathrm{i}\,\hat{H}_\text{0}\,t/\hbar} + \pazocal{O}(\lambda)\nonumber \\
&= \sum\nolimits_{i\in\mathbf{R}_\alpha} e^{-\mathrm{i}\,\omega_i\,t}\hat{A}_i + \pazocal{O}(\lambda).
\label{eq:X_int_Floquet_weak}
\end{align}
This is due to the fact that $ \hat{H}_\text{d} $ is $ \pazocal{O}(\lambda) $ while $ \hat{H}_\text{0} $ is $ \pazocal{O}(1) $. Exploiting Eq.~\eqref{eq:X_int_Floquet_weak} and following the exact same standard steps that lead to Eq.~\eqref{eq:LME}, one can easily see that the master equation
\begin{equation}
\label{eq:LME_local}
\frac{\diff \hat{\rho}_\text{s}}{\diff t} = -\frac{\mathrm{i}}{\hbar}[\hat{H}_\text{0} + \hat{H}_\text{d}(t),\,\hat{\rho}_\text{s}]+\left(\pazocal{L}_\text{c} + \pazocal{L}_\text{h}\right)\,\hat{\rho}_\text{s}
\end{equation}
would hold up to $ \pazocal{O}(\lambda\max\{\gamma_c,\gamma_h\}) $. 

Effectively, Eq.~\eqref{eq:LME_local} assumes that the dissipation is entirely decoupled from the intrinsic dynamics of $ S $, which includes the driving. This is reminiscent of the `local' master equations which are customarily used when dealing with weakly interacting multipartite open quantum systems \cite{PhysRevE.76.031115,gonzalez2017testing,Hofer2017}. We want to emphasize that, just like we have done here, \textit{it is very important to establish precisely the range of validity of such local equations} \cite{trushechkin2016perturbative} since using them inconsistently can lead to violations of the laws of thermodynamics \cite{levy2014local,stockburger2016thermodynamic}. 

It is convenient to move into the rotating frame $ \hat{\rho}_\text{s} \mapsto e^{\mathrm{i}\,\hat{H}_\text{0}\,t/\hbar}\hat{\rho}_\text{s}e^{-\mathrm{i}\,\hat{H}_\text{0}\,t/\hbar}\coloneqq \hat{\sigma}_\text{s} $ in order to remove the explicit time dependence on the right-hand side of Eq.~\eqref{eq:LME_local} and simplify the calculations. This gives 
\begin{equation} \label{eq:LMERF}
\frac{\diff \hat{\sigma}_\text{s}}{\diff t} = -\frac{\mathrm{i}}{\hbar}[\hat{h}_\text{d},\, \hat{\sigma}_\text{s}] + \left(\pazocal{L}_{c} + \pazocal{L}_{h}\right)\hat{\sigma}_\text{s},
\end{equation}
where the Hamiltonian $ \hat{H}_\text{d}(t) $ in the rotating frame is given by
\begin{equation} \label{eq:HSSwR}
\hat{h}_\text{d} \simeq \hbar\lambda \left(\ket{j_w}\bra{j_w+1} + \text{h.c.}\right).
\end{equation}
Here, we have neglected two fast-rotating terms, with frequencies $\pm 2\, \omega_{j_w} $. This is consistent with our weak driving approximation $ \tau_\text{s--d} \gg \tau_\text{R} $, as all quantities here have been time-averaged over one period of the driving field. 

The `decay rates' $ \Gamma_\omega^{\, \alpha} $ from Eq.~\eqref{eq:dissipator} are the only missing pieces to proceed to calculate the thermodynamic variables in the non-equilibrium steady state of $ S $. These are
\begin{equation}
\Gamma_\omega^{\, \alpha} = 2 \Re\,\int_0^\infty \diff r~e^{\mathrm{i}\,\omega\,t}~{\rm Tr}\{\hat{B}_\alpha(t)\hat{B}_\alpha(t-r)\hat{\rho}_\alpha\},    
\end{equation}
where the operator $ \hat{\rho}_\alpha $ represents the thermal state of bath $ \alpha $. Assuming $ d_\alpha $-dimensional baths with an infinite cutoff frequency, the decay rates become \cite{Breuer2002}
\begin{subequations}
\begin{eqnarray}
\Gamma_\omega^{\, \alpha} &=& \gamma_\alpha\, (\omega/\omega_0)^{d_\alpha} \, \left[1-\exp\left(-\hbar\omega/k_B T_\alpha \right)\right]^{-1}, \label{eq:Gw} \\
\Gamma_{-\omega}^{\, \alpha} &=& \exp{\left(-\hbar\omega/k_B T_\alpha \right)}\,\Gamma^{\, \alpha}_\omega , \label{eq:Gmw}
\end{eqnarray} 
\end{subequations}
with $\omega>0$. In our model, $ \omega_0 $ depends on the physical realization of the system-bath coupling. $ d_\alpha = 1 $ would correspond to Ohmic dissipation and $ d_\alpha = \{2,3\} $, to the super-Ohmic case.

\subsection{Heat currents, power, and performance}\label{sec:thermo_variables_S}

As it is standard in quantum thermodynamics, we will use the master equation \eqref{eq:LMERF} to break down the average energy change of the bare system $ \frac{\diff}{\diff t}\langle E \rangle(t) = \tr\,\{\hat{H}_\text{0}\, \frac{\diff}{\diff t}\hat{\sigma}_\text{s}(t)\} $ into `heat' and `power' contributions [i.e., $ \frac{\diff}{\diff t}\langle E \rangle(t) = \sum\nolimits_\alpha \dot{\mathcal{Q}}_\alpha(t) + \mathcal{P}(t) $]. These can be defined as \cite{Uzdin2015}
\begin{subequations}
\begin{align}
\dot{\mathcal{Q}}_{\alpha}(t) &\coloneqq \tr\,\{\hat{H}_\text{0}\, \pazocal{L}_\alpha\,\hat{\sigma}_\text{s}(t)\}, \label{eq:Qa}\\
\mathcal{P}(t) &\coloneqq -\mathrm{i}\,\hbar^{-1}\tr\,\{\hat{H}_\text{0}\,[\hat{h}_\text{d}, \hat{\sigma}_\text{s}(t)]\}. \label{eq:P}
\end{align}
\label{eq:thermo_variables}
\end{subequations}
In the long-time limit, $ \frac{\diff}{\diff t}\langle E \rangle \xrightarrow[]{t\rightarrow\infty} 0 $, and we will denote the corresponding steady-state heat currents and stationary power input by $ \heat{\alpha} $ and $\pow $, respectively. In particular, from Eqs.~\eqref{eq:dissipator} and \eqref{eq:thermo_variables} it can be shown that \cite{alicki1979engine,spohn1978entropy}
\begin{subequations}
\begin{align}
\heat{c} + \heat{h} + \pow &= 0, \label{eq:first_law}\\
\frac{\heat{c}}{T_c} + \frac{\heat{h}}{T_h} &\leq 0, \label{eq:second_law}
\end{align}
\label{eq:laws}
\end{subequations}
which amount to the First and Second Law of thermodynamics. It is important to note that the strict negativity of Eq.~\eqref{eq:second_law} follows directly from the geometric properties of the dynamics generated by the local dissipators $ \pazocal{L}_\alpha $ \cite{spohn1978entropy}. Working with any other reference frame to quantify energy exchanges in Eqs.~\eqref{eq:thermo_variables}, such as, e.g., $ \hat{H}_0 + \hat{h}_\text{d} $, would lead to undesirable violations of the Second Law \cite{levy2014local}. We thus see that the choice of the eigenstates of the bare Hamiltonian as preferred basis has a sound thermodynamic justification.

Using Eqs.~\eqref{eq:H0}, \eqref{eq:dissipator}, \eqref{eq:HSSwR}, and \eqref{eq:thermo_variables}, we find for our cyclic $ N $-level model
\begin{subequations}
\begin{eqnarray}
\dot{\mathcal{Q}}_{\alpha}&=&\sum\nolimits_{i\in R_\alpha} (E_{i+1}-E_i)\, J_i \,, \label{eq:Qass} \\
\mathcal{P}&=&2\hbar\lambda\,\omega_{j_w} \,\Im\,\bra{j_w}\hat{\sigma}_\text{s}(\infty)\ket{j_w+1}  \label{eq:Pss},
\end{eqnarray}
\label{eq:thermo_variables_explicit}
\end{subequations}
where $ J_i = \Gamma_{-\omega_i}^{\,\alpha_i}\,p_i^{(\infty)} - \Gamma_{\omega_i}^{\,\alpha_i}\,p_{i+1}^{(\infty)} $ is the net stationary transition rate from $ \ket{i} $ to $ \ket{i+1} $ and $ p_i^{(\infty)} \coloneqq \bra{i\,}\hat{\sigma}_\text{s}(\infty)\ket{\,i} $. The super-index $ \alpha_i $ stands for the bath associated with the dissipative transition $ \ket{i}\leftrightarrow\ket{i+1} $. Crucially, Eq.~\eqref{eq:Pss} implies that vanishing stationary quantum coherence results in vanishing power consumption ($ \pow = 0 $) and hence, no refrigeration (see also, e.g., Ref.~\cite{Kosloff2014}). In fact, as we shall see below, $ \heat{\alpha} = 0 $ in absence of coherence. Therefore, our cyclic model in Fig.~\ref{fig:fig1} is \textit{inherently quantum} since it requires non-zero coherences to operate. 

\subsection{Steady-state populations and coherence}

The key to understand why our weakly driven cyclic devices can be emulated classically resides in the interplay between populations and coherence in the eigenbasis $\{ \ket{i} \}_{i=1}^N$ of the bare Hamiltonian $ \hat{H}_\text{0}$. In this representation, Eq.~\eqref{eq:LMERF} reads:

\begin{subequations}
\begin{align}   
&\frac{\diff p_i}{\diff t} = \Gamma_{-\omega_{i-1}}^{\, \alpha_{i-1}}p_{i-1} - \left(\Gamma_{\omega_{i-1}}^{\, \alpha_{i-1}} + \Gamma_{-\omega_{i}}^{\, \alpha_{i}}\right)\, p_i+\Gamma_{\omega_{i}}^{\, \alpha_{i}}\, p_{i+1}, \label{eq:populationsa} \\
&\frac{\diff p_{j_w}}{\diff t} = \Gamma_{-\omega_{{j_w}-1}}^{\, \alpha_{{j_w}-1}}p_{{j_w}-1}-\Gamma_{\omega_{{j_w}-1}}^{\, \alpha_{{j_w}-1}}\, p_{j_w}-2\lambda\, \Im\,\bra{{j_w}}\hat{\sigma}_\text{s} \ket{{j_w}+1}, \\
&\frac{\diff p_{{j_w}+1}}{\diff t} = \Gamma_{\omega_{{j_w}+1}}^{\, \alpha_{{j_w}+1}}\, p_{{j_w}+2}-\Gamma_{-\omega_{{j_w}+1}}^{\, \alpha_{{j_w}+1}} p_{{j_w}+1} + 2\lambda\, \Im\,\bra{{j_w}}\hat{\sigma}_\text{s} \ket{{j_w}+1},
\end{align}
\label{eq:populations}
\end{subequations}
for $ i \neq \{{j_w}, {j_w}+1\}$. Note that we have omitted the time labels in $ p_k = \bra{k} \hat{\sigma}_\text{s}(t) \ket{k} $ for brevity. Importantly, the populations of the pair of levels coupled to the driving field do depend on the coherence between them. In turn, this coherence evolves as 
\begin{multline} \label{eq:coherence}
\frac{\diff}{\diff t}\bra{{j_w}}\hat{\sigma}_\text{s} \ket{{j_w}+1} \\
= -\frac{1}{2}\left(\Gamma_{\omega_{{j_w}-1}}^{\, \alpha_{{j_w}-1}} + \Gamma_{-\omega_{{j_w}+1}}^{\, \alpha_{{j_w}+1}}\right)\, \bra{{j_w}}\hat{\sigma}_\text{s} \ket{{j_w}+1} - \mathrm{i}\lambda  \left( p_{{j_w}+1} - p_{j_w} \right).
\end{multline}
Hence, the steady-state coherence (i.e., $ \frac{\diff}{\diff t} \bra{{j_w}}\hat{\sigma}_\text{d}\ket{{j_w}+1}_\infty \coloneqq 0 $) in the subspace spanned by $ \ket{{j_w}} $ and $ \ket{{j_w}+1} $ is given in terms of the steady-state populations \textit{only}. Namely, as
\begin{equation} \label{eq:coherencess}
\bra{{j_w}} \hat{\sigma}_\text{s} \ket{{j_w}+1}_{\infty} = \mathrm{i}\frac{2\lambda \, \left(p_{{j_w}}^{(\infty)}-p_{{j_w}+1}^{(\infty)}\right)}{\Gamma_{\omega_{{j_w}-1}}^{\, \alpha_{{j_w}-1}}+\Gamma_{-\omega_{{j_w}+1}}^{\, \alpha_{{j_w}+1}}}.
\end{equation}

Inserting Eq. \eqref{eq:coherencess} in \eqref{eq:populations} and imposing $ \frac{\diff}{\diff t} p_k^{(\infty)} = 0 $ yields a linear system of equations for the $ N $ stationary populations $ \mathbf{p}_{\infty} \coloneqq \big(p_1^{(\infty)}, p_2^{(\infty)}, \hdots , p_N^{(\infty)}\big)^\mathsf{T} $. This can be cast as $ \mathbf{W}\,\mathbf{p}_{\infty} = 0 $, where the non-zero elements of the `matrix of rates' $ \mathbf{W} $ are
\begin{align}\label{eq:Wi}
\begin{split}
W_{i,\,i+1} &= \Gamma_{\omega_i}^{\, \alpha_i},\,W_{i+1,\, i} = \Gamma_{-\omega_i}^{\, \alpha_i} , \\ 
W_{i,\, i-1} &= \Gamma_{-\omega_{i-1}}^{\, \alpha_{i-1}},\,W_{i-1,\, i} = \Gamma_{\omega_{i-1}}^{\, \alpha_{i-1}} , \\ 
W_{i,\, i} &=-\, \left(\Gamma_{\omega_{i-1}}^{\, \alpha_{i-1}} + \Gamma_{-\omega_i}^{\, \alpha_i} \right) ,
\end{split}
\end{align}
for $i\neq \{{j_w}, {j_w}+1\}$, and
\begin{align} \label{eq:Wj}
\begin{split}
W_{{j_w},\, {j_w}+1} &= W_{{j_w}+1,\, {j_w}} = 4\, \lambda^2 \left[\Gamma_{\omega_{{j_w}-1}}^{\, \alpha_{{j_w}-1}}+\Gamma_{-\omega_{{j_w}+1}}^{\, \alpha_{{j_w}+1}}\right]^{-1} , \\
W_{{j_w},\,{j_w}} &= - \left(\Gamma_{\omega_{{j_w}-1}}^{\, \alpha_{{j_w}-1}} + W_{{j_w},\, {j_w}+1} \right) , \\
W_{{j_w}+1,\, {j_w}+1} &= -\left(\Gamma_{-\omega_{{j_w}+1}}^{\, \alpha_{{j_w}+1}} + W_{{j_w},\, {j_w}+1}\right) .
\end{split}
\end{align}

Note that, even if $ \mathbf{W}\,\mathbf{p}_\infty = 0 $ (together with the normalization condition) determines the stationary populations of our model, Eqs.~\eqref{eq:populations} certainly differ from
\begin{equation} \label{eq:counterpart_dynamics}
\frac{\diff \mathbf{q}(t)}{\diff t} = \mathbf{W}\,\mathbf{q}(t).
\end{equation}
That is, the $ \mathbf{q}(t) $ defined by Eq.~\eqref{eq:counterpart_dynamics} converges to $ \mathbf{p}_\infty $ asymptotically but it does not coincide with $ \big( p_1(t), \cdots p_N(t) \big)^\mathsf{T} $ at any finite time. Nonetheless, as we shall argue below, a classical system $ S' $ made up of $ N $ discrete states evolving as per Eq.~\eqref{eq:counterpart_dynamics} can emulate the steady-state energy conversion process of the quantum-coherent system $ S $. Note that a similar trick has been used in \cite{Houlebec2018}, for an absorption refrigerator model where coherences appear accidentally, due to degeneracy. In contrast, as argued in Sec.~\ref{sec:thermo_variables_S}, the quantum coherence in our model is \textit{instrumental} for its operation.

Remarkably, an equation similar to \eqref{eq:counterpart_dynamics} can be found for an arbitrary choice of basis. Crucially, however, the physically motivated choice of the eigenbasis of $\hat{H}_0$ ensures the positivity of all non-diagonal rates in $ \mathbf{W} $, that they obey detailed balance relations [cf. Eqs.~\eqref{eq:detailed_balance} below], and the structure of Eq.~\eqref{eq:Qass}. All these are necessary conditions for building a classical emulator for the energy-conversion process, as we show below.

\section{Classical emulators} \label{sec:counterpart}

\subsection{General properties}\label{sec:simulator_properties}

Let us now discuss in detail the properties of the emulator $ S' $ as defined by Eq.~\eqref{eq:counterpart_dynamics}. First of all, note that \eqref{eq:counterpart_dynamics} is a proper balance equation since: (i) the non-diagonal elements of $ \mathbf{W} $ are positive, (ii) the pairs $ \{W_{i,\, i+1},\, W_{i+1,\, i}\} $ satisfy the detailed balance relations [cf. Eqs.~\eqref{eq:Gmw}, \eqref{eq:Wi}, and \eqref{eq:Wj}]
\begin{subequations}
\begin{align} 
\frac{W_{i+1,\, i}}{W_{i,\, i+1}} &= \exp\left(-\frac{\hbar\omega_i}{k_B T_{\alpha_i}}\right), \label{eq:Wi1_Wi}\qquad (i \neq {j_w}) \\
\frac{W_{{j_w}+1,\, {j_w}}}{W_{{j_w},\, {j_w}+1}} &= 1 , \label{eq:Wj1_Wj} 
\end{align}
\label{eq:detailed_balance}
\end{subequations}
and (iii) the sum over columns in $ \mathbf{W} $ is zero (i.e., $ W_{k,\,k}=-\sum_{l \neq k} W_{l,\, k} $), reflecting the conservation of probability. This also implies that $ \mathbf{W} $ is singular and that $ \mathbf{p}_\infty $ is given by its non-vanishing off-diagonal elements.

Notice as well that rates like $ \{W_{i,\, i+1},\, W_{i+1,\, i}\}_{i\neq {j_w}} $ in Eq.~\eqref{eq:detailed_balance} can always be attributed to excitation/relaxation processes (across a gap $ \hbar\omega_i $) mediated by a heat bath at temperature $ T_{\alpha_i} $. On the contrary, $ \{W_{{j_w},\, {j_w}+1},\, W_{{j_w}+1, {j_w}}\} $ indicate saturation. Therefore, one possible physical implementation of $ S' $ would be an $N$-state quantum system connected via suitably chosen coupling strengths to a hot and a cold bath as well as to a work repository, such as an infinite-temperature heat bath. Indeed, looking back at Eqs.~\eqref{eq:Wi}, we see that the rates for the dissipative interactions in $ S' $ are identical to those in $ S $. However, according to Eq. \eqref{eq:Wj}, the coupling to the driving is much smaller in the emulator than in the original quantum-coherent model. Hence, $ S' $ is \textit{not} the result of dephasing the $ N $-level cyclic machine, but a different device.

At this point, we still need to show that the steady-state energy fluxes of our emulator actually coincide with Eqs.~\eqref{eq:thermo_variables_explicit}. To do so, we now build a classical representation of $ S' $ based on graph theory. Importantly, this also serves as the generic physical embodiment for the emulator.

\subsection{Graph representation and thermodynamic variables}\label{sec:graph_thermo_variables}

\begin{figure}[t]
\includegraphics[width=\linewidth]{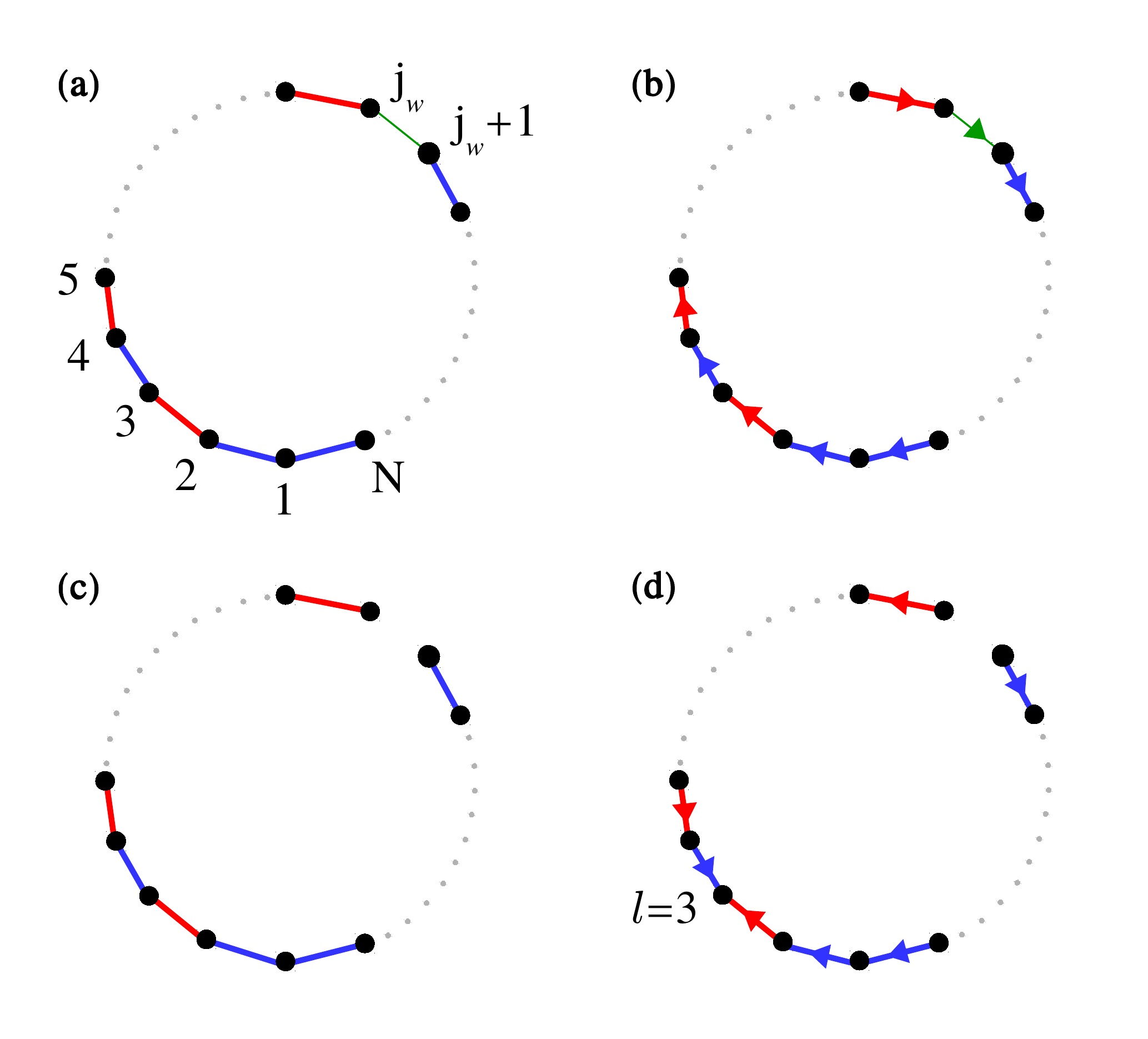}
\caption{(colour online) \textbf{(a)} The circuit graph $ \pazocal{C}_N $ is a classical emulator for the steady state of the $ N $-level quantum-coherent device of Fig.~\ref{fig:fig1}. The blue edges are associated with dissipative transitions mediated by the cold bath. Similarly, the red edges and the green edge relate to the hot bath and the external driving. \textbf{(b)} When orienting the edges of $ \pazocal{C}_N $ clockwise, the cycle $ \vec{\pazocal{C}}_N $ is obtained. \textbf{(c)} Removing the edge $({j_w},{j_w}+1) $ from $ \pazocal{C}_N $ yields the maximal tree $\pazocal{T}_{j_w}$. \textbf{(d)} Orienting $\pazocal{T}_{j_w}$ towards, e.g., vertex $l=3$ gives the oriented maximal tree $\vec{\pazocal{T}}_{j_w}^{\,l}$.} \label{fig:fig2}
\end{figure}

Using graph theory for the thermodynamic analysis of a system described by a set of rate equations has two major advantages: First, it provides a clear interpretation of the underlying energy conversion mechanisms \cite{Gonzalez2016} and, secondly, it allows for the calculation of the thermodynamic variables directly from the matrix of rates \cite{Hill1966}. In fact, the graph itself also follows from $ \mathbf{W}\ $---each state $ k = \{1,2,\hdots,N\} $ becomes a `vertex' and each pair of non-vanishing rates $ \{W_{k,\, l},\, W_{l,\, k}\}_{k\neq l} $ becomes an `undirected edge' $ (k,l) $ connecting $ k $ and $ l $. Specifically, this mapping would take Eq.~\eqref{eq:counterpart_dynamics} into the `circuit graph' $ \pazocal{C}_N $ depicted in Fig.~\ref{fig:fig2}(a). We can thus think of $ S' $ as a classical device transitioning cyclically between $ N $ states. 

Now that we have a physical picture in mind, we show that such classical emulator is thermodynamically equivalent to the coherent $ N $-level machine, once in its steady state. Specifically, we focus on rewriting the thermodynamic variables of $ S $ [cf. Eqs.~\eqref{eq:thermo_variables_explicit}] in terms of elements of $ S' $. To that end, some graph objects need to be introduced. For instance, the undirected graph $ \pazocal{C}_N $ may be oriented (or directed) either clockwise or anticlockwise, leading to the `cycles' $ \vec{\pazocal{C}}_N $ and $ -\vec{\pazocal{C}}_N $, respectively [see Fig. \ref{fig:fig2}(b)]. We may want to eliminate an edge from $ \pazocal{C}_N $, e.g., $ (k,k+1) $; the resulting undirected graph would be the `maximal tree' $ \pazocal{T}_k $ [see Fig.~\ref{fig:fig2}(c)]. Maximal trees can also be oriented towards a vertex, e.g., $ l $; the corresponding graph would then be denoted by $ \vec{\pazocal{T}}_k^{\,l} $ [see Fig.~\ref{fig:fig2}(d)]. Finally, if the edge $ (k,l) $ is directed, e.g., from vertex $ k $ to vertex $ l $, we may pair it with the transition rate $ W_{l,k} $. Similarly, any directed subgraph $ \vec{\pazocal{G}} $, {\color{myblue} for example the cycle $ \vec{\pazocal{C}}_N $ or the oriented maximal tree $ \vec{\pazocal{T}}_k^{\,l} $}, can be assigned a numeric value $ \pazocal{A}(\vec{\pazocal{G}}) $ given by the product of the transition rates of its directed edges. {\color{myblue} In particular}, $ \pazocal{A}(\vec{\pazocal{C}}_N) = \Pi_{n=1}^N W_{n+1,\, n} $, $ \pazocal{A}(-\vec{\pazocal{C}}_N)=\Pi_{n=1}^N W_{n,\, n+1} $ and
\begin{equation}
\pazocal{A}(\vec{\pazocal{T}}_k^{\,l} )= 
\left\{ \begin{array}{lcc}
\Pi_{n=1}^{l-1} W_{n+1,\, n} \ \Pi_{n=l+1}^{k} W_{n-1,\, n} \ \Pi_{n=k+1}^{N} W_{n+1,\, n} & l < k \\ \\
\Pi_{n=1}^{k-1} W_{n,\, n+1} \ \Pi_{n=k+1}^{l-1} W_{n+1,\, n} \ \Pi_{n=l}^{N} W_{n,\, n+1}  &  l > k+1  \\ \\
\Pi_{n=1}^{k-1} W_{n+1,\, n} \ \Pi_{n=k+1}^{N} W_{n+1,\, n}  &  l=k  \\ \\
\Pi_{n=1}^{k-1} W_{n,\, n+1} \ \Pi_{n=k+1}^{N} W_{n,\, n+1}  &  l=k+1  \\
\end{array}
\right.
\end{equation}\label{eq:A(T)}
where {\color{myblue}$W_{N,\, N+1}\equiv W_{N,\, 1}$ and $W_{N+1,\, N}\equiv W_{1,\, N}$}.

Our aim is to cast $ \heat{\alpha} $ and $ \pow $ solely as functions of graph objects referring to $ \pazocal{C}_N $. Let us start by noting that, the steady state populations $ p_i^{(\infty)} $ of $ S $ can be written as \cite{Schnakenberg1976}
\begin{equation} \label{eq:piss}
p_i^{(\infty)} = \pazocal{D}(\pazocal{C}_N)^{-1} \sum\nolimits_{k=1}^N\pazocal{A}(\vec{\pazocal{T}}_k^{\,i}), 
\end{equation}
since, by definition, they coincide with those of $ S' $. Here, $\pazocal{D}(\pazocal{C}_N) = \sum\nolimits_{k=1}^N \sum\nolimits_{l=1}^N \pazocal{A}(\vec{\pazocal{T}}_k^{\,l})$. Introducing Eq.~\eqref{eq:piss} in the definition of $ J_i $ [see text below Eqs.~\eqref{eq:thermo_variables_explicit}], we get
\begin{equation} \label{eq:Ji}
J_i = \pazocal{D}(\pazocal{C}_N)^{-1} \sum\nolimits_{k=1}^N \left[ W_{i+1,\, i}\, \pazocal{A}(\vec{\pazocal{T}}_k^{\,i})-W_{i,\, i+1}\,\pazocal{A}(\vec{\pazocal{T}}_k^{\,i+1})\right].
\end{equation}
The bracketed term in Eq.~\eqref{eq:Ji} turns out to be $\big[\pazocal{A}(\vec{\pazocal{C}}_N)-\pazocal{A}(-\vec{\pazocal{C}}_N)\big]\, \delta_{ki} $ \cite{Gonzalez2017}, where $\delta_{ki}$ stands for the Kronecker delta. Therefore $ J_i $ does not depend on $ i $
\begin{equation}\label{eq:Ji2}
J_i = \pazocal{D}(\pazocal{C}_N)^{-1} \left[\pazocal{A}(\vec{\pazocal{C}}_N)-\pazocal{A}(-\vec{\pazocal{C}}_N)\right] \coloneqq J.
\end{equation}
That is, in the steady state, the system exchanges energy with both baths and the driving field, {\color{myblue} with the same flux \cite{Schnakenberg1976}}. This `tight-coupling' condition between thermodynamic fluxes \cite{Esposito2009} implies that our $ N $-level device is `endoreversible' \cite{Correa2014b} and hence, that it can operate in the reversible limit of maximum energy-efficiency \cite{Correa2014a}. 

As a result, Eq.~\eqref{eq:Qass} becomes $ \heat{\alpha} = J \sum\nolimits_{i\in\mathbf{R}_\alpha} (E_{i+1}-E_i) $. Using \eqref{eq:Wi} and \eqref{eq:Wi1_Wi}, we can see that
\begin{equation} \label{eq:Xa}
\sum\nolimits_{i\in R_\alpha} (E_{i+1}-E_i) = -T_\alpha k_B \ln{\frac{\pazocal{A}^\alpha(\vec{\pazocal{C}}_N)}{\pazocal{A}^\alpha(-\vec{\pazocal{C}}_N)}} \coloneqq -T_\alpha \pazocal{X}^{\alpha}(\vec{\pazocal{C}}_N),
\end{equation}
where $ \pazocal{A}^\alpha(\pm\vec{\pazocal{C}}_N)  $ is the product of the rates of the directed edges in $ \pm\vec{\pazocal{C}}_N $ associated with bath $ \alpha $ only. Combining Eqs.~\eqref{eq:Ji2} and \eqref{eq:Xa}, we can finally express the steady-state heat currents $ \heat{\alpha} $ of the quantum-coherent $ N $-level device $ S $ as:
\begin{equation} \label{eq:Qa2}
\dot{\mathcal{Q}}_\alpha = \frac{-T_\alpha \pazocal{X}^{\alpha}(\vec{\pazocal{C}}_N)}{\pazocal{D}(\pazocal{C}_N)} \left[\pazocal{A}(\vec{\pazocal{C}}_N)-\pazocal{A}(-\vec{\pazocal{C}}_N)\right]\equiv\heat{\alpha}(\pazocal{C}_N).
\end{equation}
On the other hand, the power is easily calculated from energy conservation [cf. Eq.~\eqref{eq:first_law}]. Remarkably, the right-hand side of Eq.~\eqref{eq:Qa2} coincides with the steady-state heat currents of the circuit graph in Fig.~\ref{fig:fig2}(a), i.e., $ \heat{\alpha} \equiv \heat{\alpha}(\pazocal{C}_N) $ \cite{Hill1966}. Note that this is far from trivial, since \eqref{eq:Qa2} refers to $ S $, even if written in terms of graph objects related to $ S' $. Therefore, we have shown that the $N$-level refrigerator and its classical emulator exhibit the same stationary heat currents and power consumption and are thus, \textit{thermodynamically indistinguishable}. This is our main result. Note that the significance of Eq.~\eqref{eq:Qa2} is not just that our model is classically emulable, but that \textit{it is classically emulable in spite of requiring quantum coherence to operate} (cf. Sec.~\ref{sec:thermo_variables_S}).

\subsection{Performance optimization of the thermal machine}

The graph-theoretic expression \eqref{eq:Qa2} for the circuit currents of the classical emulator can also prove useful in the performance optimization of our quantum-coherent thermal machines $ S $ \cite{Gonzalez2017}. For instance, from the bracketed factor we see that the \textit{asymmetry} in the stationary rates associated with opposite cycles is crucial in increasing the energy-conversion rate. On the other hand, the number of positive terms in the denominator scales as $ \pazocal{D}(\pazocal{C}_N) \sim N^2 $, leading to vanishing currents. Therefore, larger energy-conversion rates are generally obtained in small few-level devices \cite{Correa2014c} featuring the largest possible asymmetry between opposite cycles.  

Let us now focus on the refrigerator operation mode (i.e., $ \dot{\mathcal{Q}}_c > 0 $, $ \dot{\mathcal{Q}}_h < 0 $, and $ \mathcal{P} > 0 $). Besides maximizing the cooling rate $ \heat{c} $, it is of practical interest to operate at large `coefficient of performance' (COP) $ \pazocal{E} $, i.e., at large cooling per unit of supplied power. In particular, the COP writes as
\begin{equation}\label{eq:cop}
\pazocal{E} \coloneqq \pazocal{E}\,(\pazocal{C}_N) = \frac{\dot{\mathcal{Q}}_c(\pazocal{C}_N)}{\mathcal{P}(\pazocal{C}_N)} = \frac{-T_c \pazocal{X}^{c}(\vec{\pazocal{C}}_N)}{T_c \pazocal{X}^{c}(\vec{\pazocal{C}}_N)+T_h \pazocal{X}^{h}(\vec{\pazocal{C}}_N)}.
\end{equation}
As a consequence of the second law [cf. Eq.~\eqref{eq:second_law}], $ \pazocal{E}(\pazocal{C}_N) $ is upper bounded by the Carnot COP ($ \pazocal{E}_C $)
\begin{equation}\label{eq:copC}
\pazocal{E} \, (\pazocal{C}_N) \leq \pazocal{E}_C = \frac{T_c}{T_h-T_c}.
\end{equation}
This limit would be saturated when $ \pazocal{X}^{h}(\pazocal{C}_N)=-\pazocal{X}^{c}(\pazocal{C}_N) $ \footnote{In Sec.~\ref{sec:graph_thermo_variables} we noted that a quantum thermal machine obeying Eqs.~\eqref{eq:populations} and \eqref{eq:coherence} is endoreversible and, therefore, capable of operating at the reversible limit of Eq.~\eqref{eq:copC}. Recall, however, that the underlying quantum master equation \eqref{eq:LME_local} is based on a \textit{local approximation}. A more accurate master equation---non-perturbative in the driving strength---can, nevertheless, be obtained using Floquet theory \cite{1205.4552v1,szczygielski2014application,Correa2014b}. Importantly, this would introduce internal dissipation \cite{PhysRevE.92.032136} neglected in Eq.~\eqref{eq:LME_local}, thus keeping the refrigerator from ever becoming Carnot-efficient \cite{Correa2014a,Correa2014b}.}. 

Since the coupling to the driving field sets the smallest energy scale in our problem, the corresponding transition rate normally satisfies $ W_{{j_w},\, {j_w}+1} \ll W_{i, \, i+1} $, $ \forall i \neq {j_w} $; unless some $ \omega_i $ becomes very small. In turn, recalling the definition of an oriented maximal tree $ \vec{\pazocal{T}}_k^l $ [see Fig.~\ref{fig:fig2}(d)], this implies $ \big\{\pazocal{A}(\vec{\pazocal{T}}_k^{\,l})\big\}_{k\neq {j_w}} \ll \big\{\pazocal{A}(\vec{\pazocal{T}}_{j_w}^{\,l})\big\}_{l=1}^N $, since the latter do not contain the small factors $ W_{{j_w},\, {j_w}+1} = W_{{j_w}+1,\, {j_w}} $, and allows for a convenient simplification of $ \pazocal{D}(\pazocal{C}_N) $ that we shall use below. Namely,
\begin{multline} \label{eq:D}
\pazocal{D}(\pazocal{C}_N)= 
\left(\frac{\sum_{k\neq {j_w}}^{N}\sum\nolimits_{l=1}^{N} \pazocal{A}(\vec{\pazocal{T}}_k^{\,l})}{\sum\nolimits_{l=1}^{N} \pazocal{A}(\vec{\pazocal{T}}_{j_w}^{\,l})} +1 \right)\, \sum\nolimits_{l=1}^{N} \pazocal{A}(\vec{\pazocal{T}}_{j_w}^{\,l})  \\ \simeq \sum\nolimits_{l=1}^N\pazocal{A}(\vec{\pazocal{T}}_{j_w}^{\,l}). 
\end{multline}

\section{Example: Power enhancement in a coherent four-level hybrid refrigerator} \label{sec:3&4leve}

In this section we apply the above to a concrete example. Namely, we solve for the steady state of two models of quantum refrigerator and find that one of them is more energy-efficient and cools at a larger rate than the other \textit{provided its steady-state coherence is also larger}. Moreover, the quantitative improvement in the cooling rate matches exactly the increase in steady-state coherence. As suggestive as this observation may seem, we then move on to show that quantum coherence is not indispensable to achieve such performance enhancement. We do so precisely by building classical emulators for both quantum-coherent models and observing that the exact same performance enhancement is possible within a fully classical stochastic-thermodynamic picture.  

\begin{figure*}[t]
\includegraphics[width=0.8\linewidth]{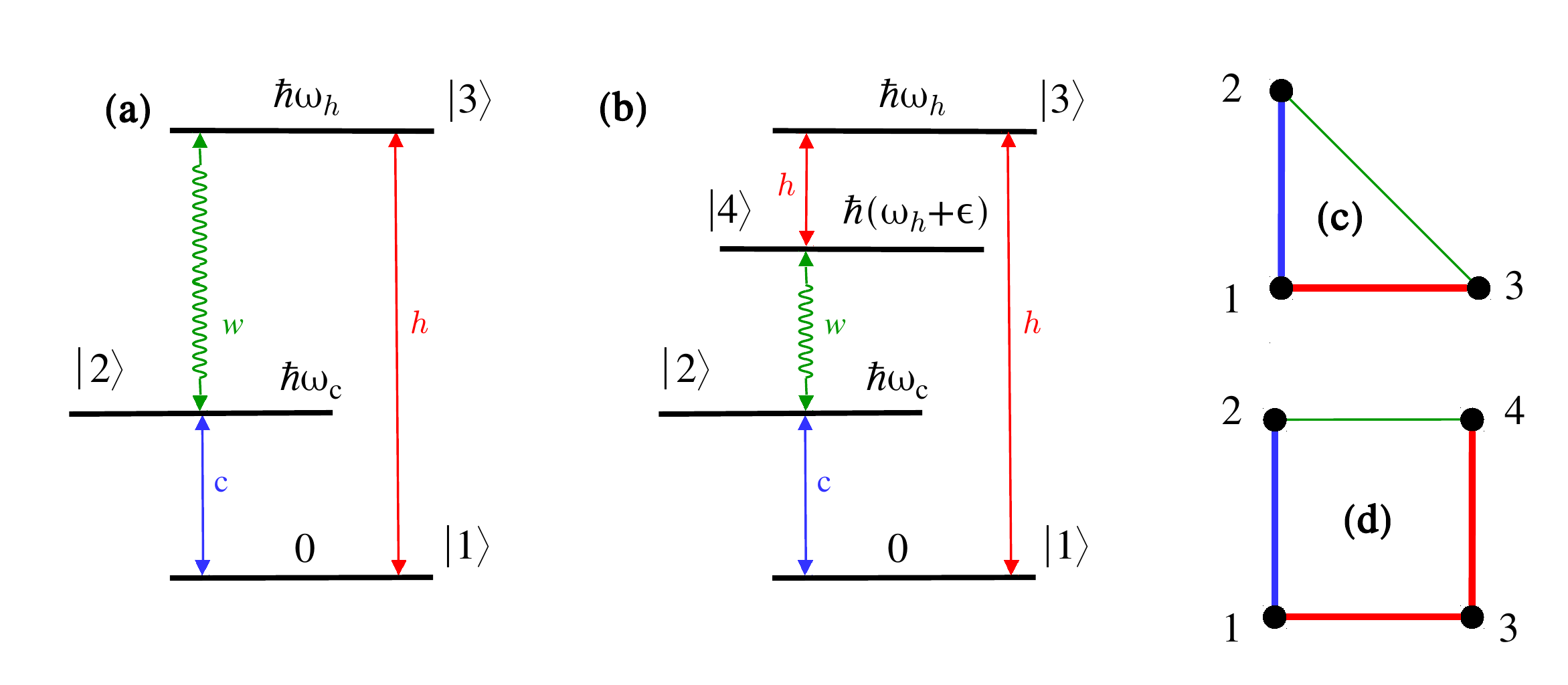}
\caption{(colour online) \textbf{(a)} \textbf{Three-level model.} The system $ S $ consists of three levels with energies $\{0,\, \hbar \omega_c,\, \hbar \omega_h\}$, such that $ 0 \leq \omega_c \leq \omega_h $. The dissipative transitions tagged $ c $ and $ h $ are mediated by a cold (blue arrow) and a hot bath (red arrow), respectively. The transition $ w $ is driven by an external sinusoidal field (curly green arrow). \textbf{(b)} \textbf{Four-level model.} An extra level with energy $ \hbar(\omega_h + \epsilon) $ is added to the three-level scheme. Note that $ \epsilon $ can be positive or negative. The driving is now applied to the transition $ \ket{2}\leftrightarrow\ket{4} $ and the new dissipative transition $ \ket{4}\leftrightarrow\ket{3} $, due to the hot bath, is added to close the thermodynamic cooling cycle. The circuit graphs associated with the three and four-level models are depicted in panels \textbf{(c)} and \textbf{(d)}, respectively. The thick blue (red) lines stand for dissipative transitions via the cold (hot) baths and the thin green lines, for the coupling to the driving.} \label{fig:fig3}
\end{figure*}
\subsection{Performance advantage}

Let us start by considering the three-level model depicted in Fig.~\ref{fig:fig3}(a). As we can see, this simple design consists of three states $ \{\ket{1},\ket{2},\ket{3}\} $, with energies $ \{ 0,\hbar\omega_c,\hbar\omega_h \} $ connected through dissipative interactions with a cold and a hot bath (transitions $ \ket{1}\leftrightarrow\ket{2} $ and $ \ket{1}\leftrightarrow\ket{3} $, respectively), and the action of a weak driving field (transition $ \ket{2}\leftrightarrow\ket{3} $). 

It is then straightforward to find the steady-state $ \hat{\sigma}_\text{s}(\infty) $ for the corresponding master equation \eqref{eq:LMERF}, and use Eqs.~\eqref{eq:thermo_variables} to compute the stationary heat currents $ \heat{\alpha}^{\,(3)} $ and power consumption $ \mathcal{P}^{\,(3)} $. It can be seen that the tight-coupling condition introduced in Sec.~\ref{sec:graph_thermo_variables} also applies to this case so that the coefficient of performance writes as
\begin{equation}
\pazocal{E}_3 \coloneqq \frac{\heat{c}^{\,(3)}}{\pow^{\,(3)}} = \frac{\omega_c}{\omega_w}, 
\label{eq:COP_3level}
\end{equation}
where $ \omega_w\coloneqq \omega_h-\omega_c $. 

We note that, when operating as a refrigerator, the three-level device uses the hot bath as a mere entropy sink; i.e., any excess heat is simply dumped into the hot bath and never reused. Interestingly, in order to improve the COP of actual (absorption) refrigerators it is commonplace to harness regenerative heat exchange in double-stage configurations. These recover waste heat from condensation to increase the evaporation rate of refrigerant \cite{Gordon2000}. Taking inspiration from thermal engineering, we thus add a fourth level $ \ket{4} $ with energy $ \hbar(\omega_h + \epsilon) $ as a stepping stone between $ \ket{2} $ and $ \ket{3} $. As shown in Fig.~\ref{fig:fig3}(b), we propose to use the external field only to drive the transition $ \ket{2} \leftrightarrow \ket{4} $, with gap $ \hbar(\omega_w + \epsilon) $. In order to close the thermodynamic cooling cycle, we put the hot bath to good use and connect dissipatively levels $ \ket{4} $ and $ \ket{3} $. This results in another tightly-coupled quantum refrigerator, with COP
\begin{equation}
\pazocal{E}_4 \coloneqq \frac{\heat{c}^{\,(4)}}{\pow^{\,(4)}} = \frac{\omega_c}{\omega_w + \epsilon}. 
\label{eq:COP_4level}
\end{equation}
This is \textit{larger} than $ \pazocal{E}_3 $ whenever $ \epsilon < 0 $, as intended. For comparison, recall that quantum absorption refrigerators \cite{Palao2001} entirely replace the driving by a dissipative coupling to a third bath at temperature $ T_w > T_h $. Our combined-cycle four-level model is therefore a novel hybrid design of independent interest, as it is partly driven by power and partly, by recovered waste heat.

\subsection{Power enhancement and quantum coherence}\label{sec:power_coherence}

Even if the combined-cycle four-level refrigerator is more energy-efficient than its power-driven three-level counterpart, we do not know yet whether it can also cool at a faster rate. To see this, let us define the figure of merit $ \mathcal{R} \coloneqq \heat{c}^{\,(4)}/\heat{c}^{\,(3)} $. From Eqs.~\eqref{eq:first_law} and \eqref{eq:thermo_variables_explicit}, it follows that
\begin{equation} \label{eq:Q4_Q3}
\mathcal{R} = \frac{C_{l_1}\left[\hat{\sigma}_\text{s}^{(4)}(\infty)\right]}{C_{l_1}\left[\hat{\sigma}_\text{s}^{(3)}(\infty)\right]},
\end{equation} 
where $ C_{l_1}\big[\hat{\sigma}_\text{s}^{(N)}(\infty)\big]=\Im\,\bra{j_w}\hat{\sigma}_\text{s}^{(N)}(\infty)\ket{j_w+1} $ stands for the \textit{$ l_1 $-norm of coherence} \cite{Baumgratz2014,Streltsov2017} in the stationary state $ \hat{\sigma}_\text{s}^{(N)}(\infty) $ of the $ N $-level thermal machine. This is a \textit{bona fide} quantifier of the amount of coherence involved in the steady-state operation of the device. An enhancement in the cooling rate translates into $ \mathcal{R} > 1 $ and hence, is only possible if the steady state $ \hat{\sigma}_\text{s}^{(4)}(\infty) $ of the four level device contains more quantum coherence than $ \hat{\sigma}_\text{s}^{(3)}(\infty) $. As it turns out, it is rather easy to find parameter ranges in which $ \mathcal{R} > 1 $, as shown in Figs.~\ref{fig:fig4}(a) and (b). Importantly, it is even possible to find parameters for which $ \pazocal{E}_4 > \pazocal{E}_3 $ and $ \heat{c}^{\,(4)} > \heat{c}^{\,(3)} $ \textit{simultaneously}.

\subsection{Power enhancement without quantum coherence}

Both the three and four-level refrigerators are cyclic non-degenerate heat devices and hence, classically emulable. In particular, the emulator for the three-level model is the triangle $ \pazocal{C}_3 $ depicted in Fig.~\ref{fig:fig3}(c) while the steady state of the hybrid four-level refrigerator is emulated by the square graph $ \pazocal{C}_4 $ of Fig.~\ref{fig:fig3}(d). In spite of the fact that there exist quantum coherent implementations of these energy-conversion cycles for which $ \mathcal{R} > 1 \Leftrightarrow C_{l_1}\left[\hat{\sigma}_\text{s}^{(4)}(\infty)\right] > C_{l_1}\left[\hat{\sigma}_\text{s}^{(3)}(\infty)\right] $, it would be wrong to claim that quantumness is \textit{necessary} for such performance boost---the corresponding emulators also satisfy $ \heat{c}(\pazocal{C}_4) = \heat{c}^{\,(4)} > \heat{c}(\pazocal{C}_3) = \heat{c}^{\,(3)} $, and yet have no coherence.

\subsection{Analytical insights from graph theory}

\subsubsection{Cooling rate}

Using Eqs.~\eqref{eq:Wi1_Wi}, \eqref{eq:Xa}--\eqref{eq:cop}, we readily find that
\begin{subequations}
\begin{eqnarray}
\dot{\mathcal{Q}}_c(\pazocal{C}_3)&=&\hbar\omega_c \frac{\pazocal{A}(\vec{\pazocal{C}}_3)}{\pazocal{D}(\pazocal{C}_3)} \left[1-\exp{\left(X_c-X_h\right)}\right].  \label{eq:Qc3L} \\
\dot{\mathcal{Q}}_c(\pazocal{C}_4)&=&\hbar\omega_c \frac{\pazocal{A}(\vec{\pazocal{C}}_4)}{\pazocal{D}(\pazocal{C}_4)} \left[1-\exp{\left(X_c-X_h-X_\epsilon\right)}\right].  \label{eq:Qc4L}
\end{eqnarray}
\end{subequations}
The quantities $ X_c \coloneqq \hbar \omega_c/k_B T_c $, $ X_h \coloneqq \hbar \omega_h/ k_B T_h  $, and $ X_\epsilon\coloneqq\hbar\epsilon/k_B T_h $ are `thermodynamic forces' associated with the cold and hot baths. Note that their difference encodes the \textit{asymmetry} between the two possible orientations of the graphs.

\begin{figure*}[t]
\includegraphics[width=0.3\linewidth]{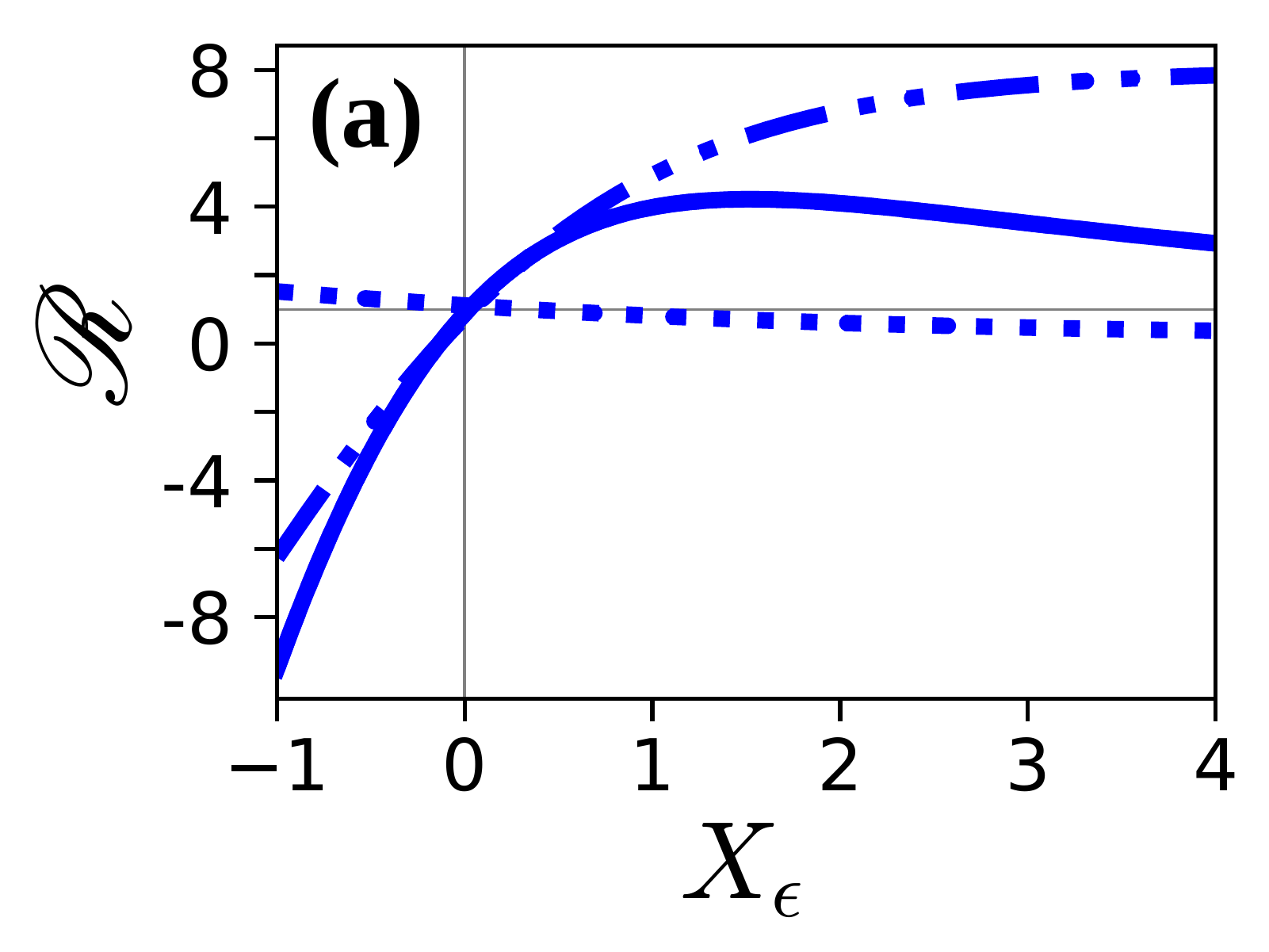}
\includegraphics[width=0.3\linewidth]{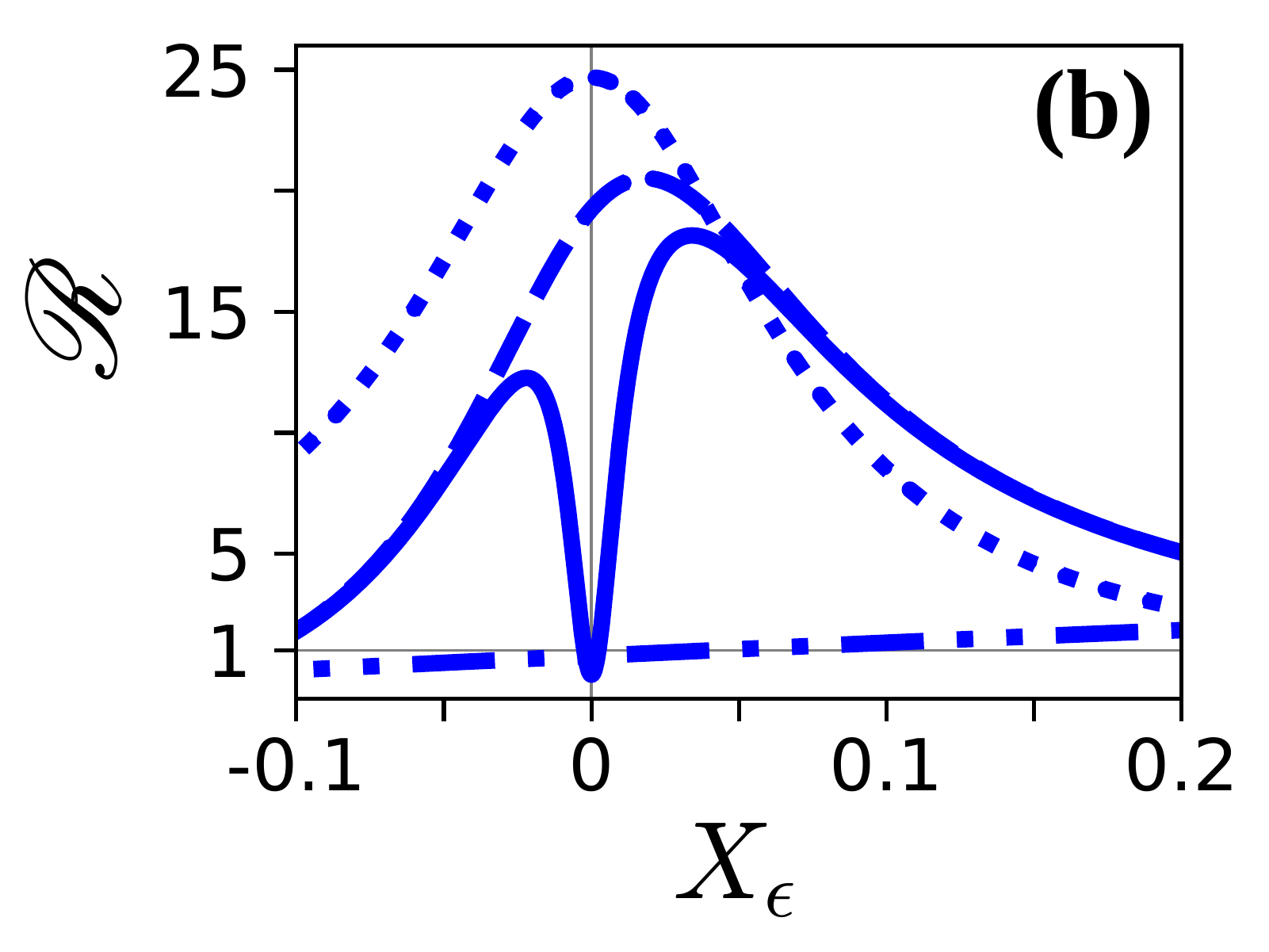}
\includegraphics[width=0.3\linewidth]{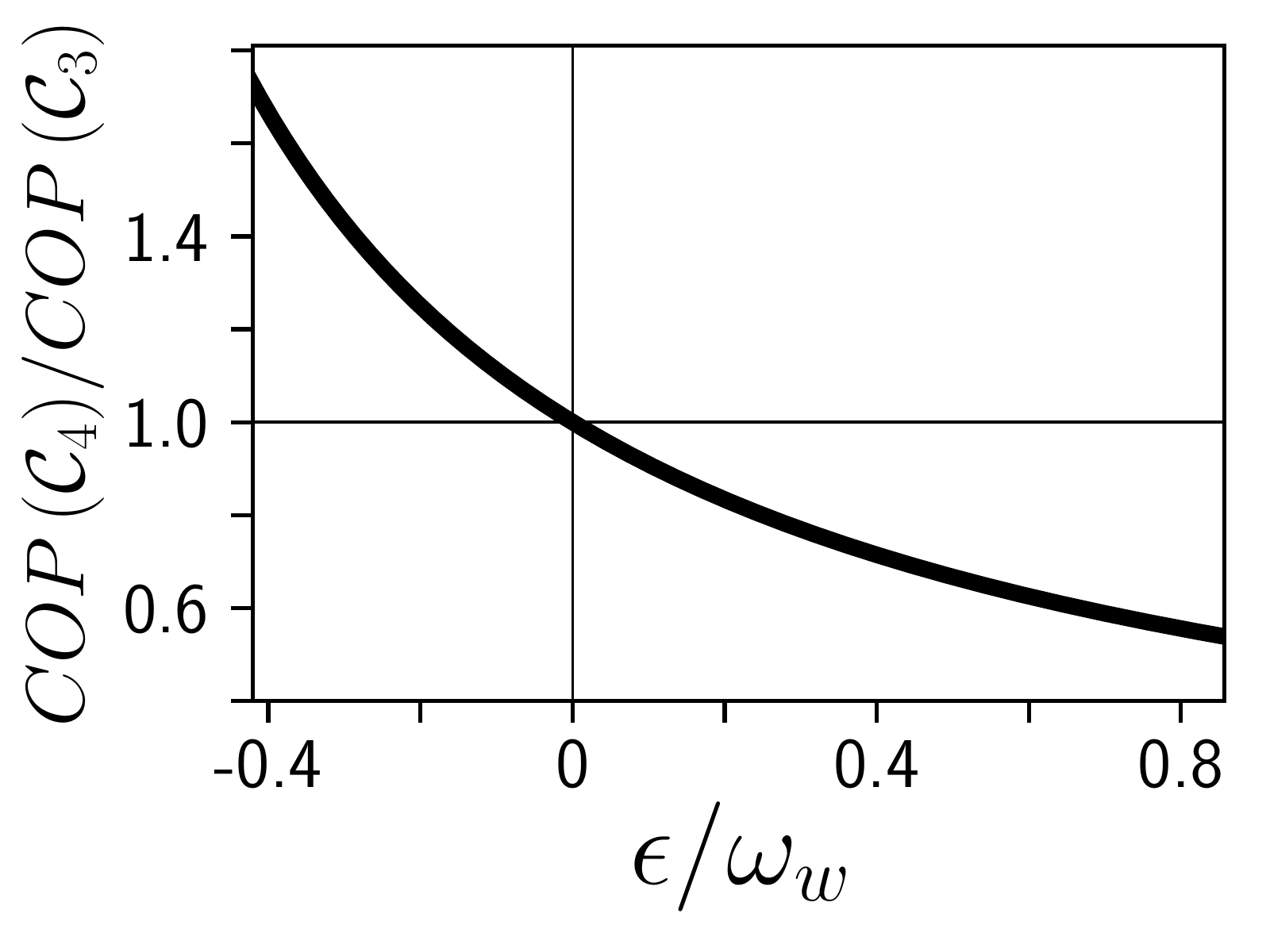}
\caption{(colour online) \textbf{(a)} Performance ratio $\mathcal{R} = \dot{\mathcal{Q}}_c(\pazocal{C}_4)/\dot{\mathcal{Q}}_c(\pazocal{C}_3)$ (solid line) as a function of the thermodynamic force $ X_\epsilon $. We also plot the approximation $ A D $ from Eq.~\eqref{eq:ratio} (dashed line), and the factors $ A $ (dotted line) and $ D $ (dot-dashed line). In this case, {\color{myblue} the solid and dashed lines are indistinguishable at the scale of the figure}. We have chosen one-dimensional baths $ d_c = d_h = 1 $, $ \omega_h = 1 $, $ \omega_c = 0.3 $, $ \lambda = 10^{-8} $, $ \gamma_c = \gamma_h = 10^{-6} $, $ T_c=1.5 $, and $ T_h=3 $ ($ \hbar = k_B = 1 $). Importantly, as discussed in Sec.~\ref{sec:power_coherence}, $ \mathcal{R} $ coincides with the ratio of stationary coherence of the four-level model and the benchmark, measured by the $ l_1 $-norm. \textbf{(b)} Same as in (a) for three-dimensional baths ($ d_c = d_h = 3 $). The rest of parameters remain unchanged. \textbf{(c)} Ratio between the COPs of $ \pazocal{C}_3 $ and $ \pazocal{C}_4 $ versus $ \epsilon $. Note that, for $ \epsilon < 0 $, the four-level device can simultaneously achieve larger cooling power \textit{and} COP. All parameters are the same as in (b).}\label{fig:fig4}
\end{figure*}

\subsubsection{Performance enhancement}

A manageable analytical approximation for the figure of merit $\mathcal{R} = \heat{c}(\pazocal{C}_4)/\heat{c}(\pazocal{C}_3)$ can be obtained by combining Eqs.~\eqref{eq:Wi}, \eqref{eq:Wj}, \eqref{eq:Qc3L}, and \eqref{eq:Qc4L} with the assumption that the smallest transition rate is the one related to the driving field [i.e., Eq.~\eqref{eq:D}]. This gives
\begin{subequations}
\begin{align}
\mathcal{R} &\simeq A D, \label{eq:ratio} \\
A &\coloneqq \frac{1+\exp{\left(X_c\right)}+\exp{\left(X_c-X_h\right)}}{1+\exp{\left(X_c\right)}+\exp{\left(X_c-X_h\right)}+\exp{\left(X_c-X_h-X_\epsilon\right)}} \nonumber\\  
&\qquad\qquad\qquad\qquad\qquad\times\frac{1-\exp{\left(X_c-X_h-X_\epsilon\right)}}{1-\exp{\left(X_c-X_h\right)}}, \\
D &\coloneqq \frac{\Gamma_{\omega_c}^c+\Gamma_{\omega_h}^h}{\Gamma_{\omega_c}^c+\Gamma_{\epsilon}^h}. \label{eq:R_ratio}
\end{align}
\label{eq:approximation}
\end{subequations}
As we can see, the factor $ A $ depends exclusively on the thermodynamic forces $ X_c $, $ X_h $ and $ X_\epsilon $. {\color{myblue} Its second term describes the ratio between the cycles asymmetries. The new thermodynamic force allows for an additional control on the asymmetry of the four-level refrigerator, thus favouring the cooling cycle for $ X_\epsilon>0 $.} In contrast, the quantity $ D $ is purely dissipative; {\color{myblue} it encodes the ratio between the rates associated with the driving fields Eq.~\eqref{eq:Wj}. Importantly, small $ \epsilon $ in the four-level model can increase $ D $. The choice of spectral density for the system-bath interactions (i.e., the dimensionality or `Ohmicity' of the baths) can lead to a sufficiently large $ D $ so that the product $ \mathcal{R}=AD > 1 $ for $ \epsilon < 0 $. Thus, factorizing $ \mathcal{R} $ as in Eqs.~\eqref{eq:approximation} provides insights into the competing physical mechanisms responsible for the cooling power enhancements in the four-level device}.

Fig.~\ref{fig:fig4}(a) illustrates how the approximation \eqref{eq:ratio} may hold almost exactly; namely, we work with one-dimensional baths ($ d_c = d_h = 1 $) at moderate to large temperatures ($ k_B T_\alpha/\hbar\omega_\alpha \gtrsim 1 $). In this range of parameters, $ A $ is the main contribution to the enhancement of $ \heat{c} $, so that $ \mathcal{R} $ is nearly insensitive to changes in the dissipation strengths. 

On the contrary, when taking three-dimensional baths ($ d_c = d_h = 3 $), the frequency-dependence of the transition rates is largely accentuated. At small $ \epsilon $ the rate $ W_{2,4} $ ceases to be the smallest in $ \pazocal{C}_4 $, which invalidates Eqs.~\eqref{eq:approximation} [see Fig.~\ref{fig:fig4}(b)] \footnote{Recall that, for Eq.~\eqref{eq:LME_local} to be valid, we must have $ 2 \epsilon \gg \gamma_h $, as required by the underlying secular approximation $ \tau_{\text{0}} \ll \tau_{\text{R}} $ [cf. Eq.~\eqref{eq:time_scales}]. The solid lines in Figs.~\ref{fig:fig4} are mere guides to the eye, which smoothly interpolate between points well within the range of validity of the master equation.}. In this case, the behaviour of $ \mathcal{R} $ is dominated by $ D $, and grows almost linearly with $ \gamma_h $. Similar disagreements between $ \mathcal{R} $ and the approximate formula Eq.~\eqref{eq:approximation} can be observed in the low-temperature regime. 

As shown in Fig.~\ref{fig:fig4}(b), the hybrid four-level design can operate at much larger energy-conversion rates than the power-driven benchmark. Indeed, for the arbitrarily chosen parameters in the figure, $ \heat{c}(\pazocal{C}_4) $ can outperform $\heat{c}(\pazocal{C}_3)$ by an order of magnitude at appropriate values of $ \epsilon $. Such enhancement may be {\color{myblue} classically interpreted} solely in terms of the asymmetry $A$ and the dissipation factor $D$, without resorting to the buildup of quantum coherence. {\color{myblue} Importantly, such qualitative understanding follows directly from the classical emulability of the model, which allows us to study the emulator in place of the original quantum-coherent device.}

\subsubsection{Coefficient of performance and cooling window}

The COPs of $\pazocal{C}_3$ and $\pazocal{C}_4$ are related through
\begin{equation}
\pazocal{E} \, (\pazocal{C}_4)/\pazocal{E}\, (\pazocal{C}_3) = (1 + \epsilon/\omega_w)^{-1}.
\end{equation}
As already mentioned, $ \epsilon < 0 $ results in an increase of the energetic performance of the four-level machine due to the lower power consumption [see Fig.~\ref{fig:fig4}(c)]. Interestingly, the ratio $ \mathcal{R} $ can be larger than one for $ \epsilon < 0 $, which entails a \textit{simultaneous power and efficiency enhancement}. For instance, comparing Figs.~\ref{fig:fig4}(b) and \ref{fig:fig4}(c), we observe increased power by a factor of $ 10 $ together with a $ 10\% $ improvement in energy-efficiency.

On the other hand, the operation mode---heat engine or refrigerator---depends on the specific parameters of the models. In particular, to achieve cooling action in the three-level benchmark we must have $ \heat{c}(\pazocal{C}_3) > 0 $. According to Eq.~\eqref{eq:Qc3L}, this implies $ X_c - X_h < 0 $ or, equivalently,
\begin{equation}
\omega_c < \omega_{c,\text{rev}}\coloneqq\omega_h\,T_c / T_h.
\label{eq:window3}
\end{equation}
Taking a fixed $ \omega_h $, the range of values $ \omega_c < \omega_{c,\text{rev}} $ is thus referred-to as \textit{cooling window}. On the other hand, from Eq.~\eqref{eq:Qc4L} we can see that cooling is possible on $ \pazocal{C}_4 $ if 
\begin{equation}
\omega_c < (\omega_h+\epsilon)\,T_c / T_h.    
\end{equation}
Hence, whenever $ \epsilon > 0 $ the cooling window of the four-level model is \textit{wider} than that of the benchmark for the \textit{same} parameters $ \omega_h $, $ T_\alpha $, $ \lambda $, $ \gamma_\alpha $, and $ d_\alpha $. Conversely, if we were interested in building a quantum heat engine, a negative $ \epsilon $ [as depicted in Fig.~\ref{fig:fig3}(b)] would broaden the operation range.

\section{Conclusions}\label{sec:conclusions}

We have analyzed periodically driven thermal machines weakly coupled to an external field and characterized by a {\it cyclic} sequence of transitions. We have proposed a new approach to build fully \textit{incoherent} classical emulators for this family of \textit{quantum-coherent} heat devices, which exhibit the exact same thermodynamic operation in the long-time limit. In particular, we exploit the fact that the steady state of this type of coherent thermal machines coincides with that of some stochastic-thermodynamic model with the same number of states dissipatively connected via thermal coupling to the same heat baths, and obeying consistent rate equations.

We have then shown how the performance of a three-level quantum-coherent refrigerator may be significantly improved by driving it with a combination of waste heat and external power---both the energy-efficiency and the cooling rate can be boosted in this way. In particular, we have shown that the cooling enhancement is identical to the increase in stationary quantum coherence, when comparing our hybrid model with an equivalent benchmark solely driven by power. In spite of the close connection between the observed effects and the buildup of additional quantum coherence, we remark that these cannot be seen as unmistakable signatures of quantumness since our model belongs to the aforementioned family of ``classically emulable'' thermal machines.

In fact, the possibility to emulate clasically a quantum heat device goes far beyond the cyclic and weakly driven models discussed here. For instance, in the opposite limit of \textit{strong} periodic driving (i.e., $ \tau_\text{d} \ll \tau_\text{R} $) one can always resort to Floquet theory to map the steady state operation of the machine into a fully incoherent stochastic-thermodynamic process in some relevant rotating frame \cite{1205.4552v1,szczygielski2014application,Correa2014b}. Graph theory can be then directly applied for a complete thermodynamic analysis \cite{Gonzalez2016}. Note that this holds for \textit{any} periodically-driven model and not just for those with a cyclic transition pattern. Similarly, \textit{all} heat-driven (or `absorption') thermal machines with non-degenerate energy spectra are incoherent in their energy basis and thus, classically emulable in the weak coupling limit \cite{Gonzalez2017}. Furthermore, the equivalence between multi-stroke and continuous heat devices in the small action limit \cite{Uzdin2015} provides a means to generalize our ``classical simulability'' argument to \textit{reciprocating} quantum thermodynamic cycles.

In this paper, we have thus extended the applicability of Hill theory \cite{Hill1966,Schnakenberg1976} to enable the graph-based analysis of a whole class of quantum-coherent thermal devices. We have also put forward a novel hybrid energy conversion model of independent interest, which exploits heat recovery for improved operation. More importantly, we have neatly illustrated why extra care must be taken when linking quantum effects and enhanced thermodynamic performance. This becomes especially delicate in, e.g., biological systems \cite{engel2007evidence}, in which the details of the underlying physical model are not fully known. 

It is conceivable that {\color{myblue} the steady state} of some {\color{myblue} continuous} quantum thermal machines might not be classically emulable---for our arguments to hold, the resulting equations of motion [analogous to our Eq.~\eqref{eq:counterpart_dynamics}] must also be proper balance equations with positive transition rates and a clear interpretation in terms of probability currents. Since the positivity of the rates is model-dependent, the search for classical emulators of more complicated devices, including consecutive driven transitions and degenerate states, might pave the way towards \textit{genuinely quantum} energy-conversion processes with no classical analogue. This interesting open question will be the subject of future work. 

\acknowledgments
We gratefully acknowledge financial support by the Spanish MINECO (FIS2017-82855-P), the European Research Council (StG GQCOP No. 637352), and the US National Science Foundation under Grant No. NSF PHY1748958. JOG acknowledges an FPU fellowship from the Spanish MECD. LAC thanks the Kavli Institute for Theoretical Physics for their warm hospitality during the program ``Thermodynamics of quantum systems: Measurement, engines, and control''.

\appendix*

{\color{myblue}
\section{A model with heat leaks} \label{Sec:appendix}
We consider here the model depicted in Fig.~\ref{fig:fig1} when adding an extra hot transition between the levels $|j_w\rangle$ and $|j_w+1\rangle$. The populations of such device fulfill Eq. \eqref{eq:populationsa} and
\begin{eqnarray}
\frac{\diff p_{j_w}}{\diff t} &=& \Gamma_{-\omega_{{j_w}-1}}^{\, \alpha_{{j_w}-1}}p_{{j_w}-1}-\Gamma_{\omega_{{j_w}-1}}^{\, \alpha_{{j_w}-1}}\, p_{j_w} \nonumber \\
&&-2\lambda\, \Im\,\bra{{j_w}}\hat{\sigma}_\text{s} \ket{{j_w}+1}+J_{j_w, \, j_w+1} \,, \nonumber \\
\frac{\diff p_{{j_w}+1}}{\diff t} &=& \Gamma_{\omega_{{j_w}+1}}^{\, \alpha_{{j_w}+1}}\, p_{{j_w}+2}-\Gamma_{-\omega_{{j_w}+1}}^{\, \alpha_{{j_w}+1}} p_{{j_w}+1} \nonumber \\
&&+ 2\lambda\, \Im\,\bra{{j_w}}\hat{\sigma}_\text{s} \ket{{j_w}+1}-J_{j_w, \, j_w+1} \,,
\end{eqnarray}
where
\begin{equation}
J_{j_w, \, j_w+1}=\Gamma_{\omega_{j_w}}^{\, h}p_{j_w+1}-\Gamma_{-\omega_{j_w}}^{\, h}p_{j_w} 
\end{equation}
is the flux from the state $|j_w\rangle$ to $|j_w+1\rangle$. This new flux is the responsible for the emergence of an additional term in the non diagonal rates
\begin{eqnarray}
W_{{j_w},\, {j_w}+1} & = 4\, \lambda^2 \left[\Gamma_{\omega_{{j_w}-1}}^{\, \alpha_{{j_w}-1}}+\Gamma_{-\omega_{{j_w}+1}}^{\, \alpha_{{j_w}+1}}\right]^{-1} + \Gamma_{\omega_{j_w}}^{\, h} \\
W_{{j_w+1},\, {j_w}} & = 4\, \lambda^2 \left[\Gamma_{\omega_{{j_w}-1}}^{\, \alpha_{{j_w}-1}}+\Gamma_{-\omega_{{j_w}+1}}^{\, \alpha_{{j_w}+1}}\right]^{-1} + \Gamma_{-\omega_{j_w}}^{\, h} \,.
\end{eqnarray}
The functions $\Gamma_{\omega_{j_w}}^{\, h}$ and $\Gamma_{-\omega_{j_w}}^{\, h}$ are always positive and fulfil a detailed balance relation at temperature $T_h$ and frequency $\omega_{j_w}$. The other non diagonal rates remain the same following Eq.~\eqref{eq:Wi}. The resulting matrix of rates allows for the definition of a graph representation; therefore, a classical emulator can be assigned also to this model. 

Such emulator is defined by a graph $\pazocal{G}$ with three circuits: the original circuit $\pazocal{C}_N$, a two-edge circuit $\pazocal{C}_{2}$---with vertices $j_w$ and $j_w+1$---corresponding to power dissipation into the hot bath, and a $N$-edge circuit $\pazocal{C}_{N}'$, where the work edge is replaced by the new hot edge, related to a heat leak from the hot to the cold bath. The heat currents and power of these circuits can be obtained by following the techniques explained in \cite{Gonzalez2016} and \cite{Gonzalez2017}. The total heat currents are then the sum of the following three thermodynamically consistent contributions
\begin{eqnarray}
\heat{\alpha}(\pazocal{C}_N)&=& \frac{-T_\alpha \pazocal{X}^{\alpha}(\vec{\pazocal{C}}_N)}{\pazocal{D}(\pazocal{G})} \left[\pazocal{A}(\vec{\pazocal{C}}_N)-\pazocal{A}(-\vec{\pazocal{C}}_N)\right] \,, \nonumber \\
\heat{\alpha}(\pazocal{C}_{2})&=& \frac{-T_\alpha \pazocal{X}^{\alpha}(\vec{\pazocal{C}}_{2})\, \det(-\mathbf{W}|\pazocal{C}_{2})}{\pazocal{D}(\pazocal{G})} \left[\pazocal{A}(\vec{\pazocal{C}}_{2})-\pazocal{A}(-\vec{\pazocal{C}}_{2})\right]  \,, \nonumber \\
\heat{\alpha}(\pazocal{C}_{N}')&=& \frac{-T_\alpha \pazocal{X}^{\alpha}(\vec{\pazocal{C}}_{N}')}{\pazocal{D}(\pazocal{G})} \left[\pazocal{A}(\vec{\pazocal{C}}_{N}')-\pazocal{A}(-\vec{\pazocal{C}}_{N}')\right]  \,, 
\end{eqnarray}
where $\pazocal{D}(\pazocal{G})$ is calculated by considering all the maximal trees of the graph containing the three circuits. Besides the matrix $\mathbf{W}|\pazocal{C}_{2}$ is obtained from the matrix of rates by removing the rows and columns corresponding to the vertices of $\pazocal{C}_{2}$.}


%

\end{document}